 \documentclass[preprint2]{aastex}

\usepackage{amssymb}
\usepackage{amsmath}
\usepackage{graphicx}
\usepackage{rotating}
\usepackage{lscape}
\usepackage{array}
\def\hi{H\,{\sc i} }
\def\hii{H\,{\sc i}{\sc i} }

\def\deg{$^{\circ}$}
\def\kms{km~s$^{-1}$}
\def\msun{M$_{\odot}$}
\def\Ha{H$\sf \alpha$}
\def\micron{$\mu$m}
\def\sfrsd{$\Sigma_{SFR}$}
\def\htwosd{$\Sigma_{H_2}$}

\def\qgas{$Q_{gas}$}
\def\sgas{$S_{gas}$}
\def\QGS{$Q_{gas,*}$}

\def\msunpps{M$_{\odot}$~pc$^{-2}$}
\def\msunppc{M$_{\odot}$~pc$^{-3}$}

\long\def\symbolfootnote[#1]#2{\begingroup%
\def\thefootnote{\fnsymbol{footnote}}\footnote[#1]{#2}\endgroup}

\shorttitle{Star Formation Models for NGC~2915 and NGC~1705}
\shortauthors{Elson~et~al.}

\begin{document}


\title{Star Formation Models for the Dwarf Galaxies\\
NGC~2915 and NGC~1705}


\author{E.~C.~Elson\altaffilmark{1,2}, W.~J.~G.~de~Blok\altaffilmark{1}, and R.~C.~Kraan-Korteweg\altaffilmark{1}}
\affil{$^1$Astrophysics, Cosmology and Gravity Centre (ACGC), Department of Astronomy, University of Cape Town, Private Bag X3,
Rondebosch 7701, South Africa}
\affil{$^2$International Centre for Radio Astronomy Research, The University of Western Australia, M468, 35 Stirling Highway, Crawley, WA, 6009 Australia}

%
%




\begin{abstract}
Crucial to a quantitative understanding of galaxy evolution are the properties of the inter-stellar medium that regulate galactic-scale star formation activity.  We present here the results of  a suite of star formation models applied to the nearby blue compact dwarf galaxies NGC~2915 and NGC~1705.  Each of these galaxies has a stellar disk embedded in a much larger, essentially star-less \hi disk.  These atypical stellar morphologies allow for rigorous tests of star formation models that examine the effects on star formation of the H\,{\sc i}, stellar and dark matter mass components, as well as the kinematics of the gaseous and stellar disks.  We use far ultra-violet  and 24~\micron\ imaging from the Galaxy Evolution Explorer and the \emph{Spitzer} Infrared Nearby Galaxies Survey respectively to map the spatial distribution of the total star formation rate surface density within each galaxy.  New high-resolution \hi line observations obtained with the Australia Telescope Compact Array are used to study the distribution and dynamics of each galaxy's neutral inter-stellar medium.  The standard Toomre~Q parameter is unable to distinguish between active and non-active star forming regions, predicting the \hi disks of the dwarfs to be sub-critical.  Two-fluid instability models incorporating the stellar and dark matter components of each galaxy, in addition to the gaseous component, yield portions of the inner disk unstable.  Finally, a formalisation in which the \hi kinematics are characterised by the rotational shear of the gas produces models that very accurately match the observations.  This suggests the time available for perturbations to collapse in the presence of rotational shear to be an important factor governing galactic-scale star formation.
\end{abstract}



\section{Introduction}\label{2915_SF2_intro}
Galaxies are gravitationally bound systems of stars, gas, and dark matter.  The interaction and evolution of these mass components results in a change of the observable properties of the galaxy as a whole.  An understanding of the processes governing galaxy formation and evolution is still a major challenge to astronomers.  The goal is to quantitatively define the relationships between galactic-scale star formation activity and the properties of the inter-stellar medium (ISM) \citep[e.g.][]{wyder_et_al_2009}.  Current observing facilities allow for high-quality multi-wavelength observations of comparable resolution and sensitivity to be carried out for nearby galaxies \citep[e.g.][]{leroy_THINGS}.  These observations are used to test current theories of galactic-scale star formation.  

The cold gas component of a galaxy is its fuel for star formation.  On global as well as localised length scales the distribution of these mass components is known to correlate with the star formation activity \citep{kennicutt_1983,kennicutt_1989,kennicutt_1998, bigiel_2008,leroy_THINGS}.  In addition to the distribution of the gas, it is also the kinematics that play a role in regulating the star formation activity.  For example, while self-gravity drives gas perturbations to collapse under mutual gravity, the centrifugal forces associated with their rotational motions will impede the structure growth by counteracting the inward gravitational force \citep{Toomre_1964,kennicutt_1989}.  Similarly, rotational shear will tear adjacent gas clouds apart before they have the chance to  coalesce to form a single denser body of gas \citep{wyder_et_al_2009}.  The self-gravity of a galaxy's ISM is determined by the gravitational potential of its mass components.  The stellar potential, for example, contributes to the self-gravity of the gas, thereby making it more susceptible to gravitational collapse and hence star formation \citep[e.g.][]{rafikov_2001,leroy_THINGS}.  Star formation models are used to understand how all of these processes work \emph{together} to control the star formation in galaxies.   Both the ability and inability of these models to accurately describe the star-formation contributes to a further understanding of the complex interplay between the various ISM properties that regulate the star formation activity.

In order to compare modelling results to observations, a quantitative measure of a galaxy's star formation activity is required.  One of the most widely used tracers of high-mass star formation is the \Ha\ emission line arising from the recombination of ionised hydrogen.  This traces a galaxy's star formation over the relatively short lifetimes (a few million years) of the most massive stars.  In recent years, GALEX observations of nearby galaxies have allowed the ultra-violet~(UV) flux originating from the photospheres of high-mass O- and B-type stars to be used as a diagnostic of the directly observable star formation rate (SFR).  As \citet{lee_2009} point out, a galaxy's UV flux probes a fuller mass spectrum of massive stars, and thus measures star formation averaged over a longer $\sim 10^8$~yr time-scale.  The UV emission alone is, however, an insufficient tracer of the SFR.  Several authors (e.g.~\citealt{salim_2007} and references therein) have compared \Ha\ and UV SFRs for nearby galaxies.  The general consensus is that UV SFRs tend to be slightly lower than \Ha\ SFRs.  The discrepancy is attributed to the effects of internal dust extinction which lowers the observed UV flux.  When UV emission from high-mass stars is absorbed by dust, it is re-emitted at infrared wavelengths.  As \citet{kennicutt_1998} points out, internal dust extinction serves as one of the largest sources of systematic error in SFR measurements.  Combining a measure of a star-forming galaxy's infrared emission with a measure of its UV emission yields a measure of the total SFR that is consistent with the rate inferred from \Ha\ spectroscopic imaging.  \citet{calzetti_2007}, for a sample of 33 nearby galaxies observed with the \emph{Spitzer} MIPS instrument at 24~\micron, showed that using the 24~\micron\ emission to account for the dust extinction allows for an accurate determination of the total number of ionising photons in an HII region.  

Dwarf galaxies are particularly useful laboratories in which to study star formation.  Being morphologically and dynamically simpler systems than larger spiral galaxies, they allow us to stringently test star formation theories in the domains of low gas surface densities, extreme dark matter content, and lower rotational shear rates.  This paper presents the results of a detailed study of the star formation activity in two nearby dwarfs, NGC~2915 and NGC~1705.  Each galaxy has a small stellar disk embedded in a much larger \hi disk.  Using the Australia Telescope Compact Array we have obtained new deep, high-resolution \hi line observations of these galaxies in order to study their extended \hi disks in unprecedented detail.  We aim to link the \hi characteristics of each galaxy to its observed star formation activity.  Far ultra-violet and 24~\micron\ imaging from the GALEX and \emph{Spitzer} satellites, respectively is used to quantify the star formation activity. We combine all of these data to produce a suite of detailed star formation models that are used to understand the observed star formation in the context of the distribution and kinematics of the neutral ISM.  The results of our analyses are compared to those of \citet{leroy_THINGS} who carried out similar analyses for their sample 23 nearby late-type galaxies from The \hi Nearby Galaxy Survey \citep[THINGS,][]{THINGS_walter}.  This is done so that the results can be interpreted in the context of other more typical star-forming galaxies.  

The structure of this paper is as follows: Section~\ref{n2915_n1705} introduces NGC~2915 and NGC~1705.  The new H\,{\sc i}, ultra-violet and 24~\micron\ data sets that are utilised in this work are presented in Sec~\ref{data}.  The method of estimating the total star formation rates is presented in Sec.~\ref{2915_FUV_IR_SFR_maps_sec} together with total star formation rate maps of the galaxies.  Each of the star formation models is introduced and described in detail in Section~\ref{star_formation_models}.  The modeling results for each galaxy are also shown in Section~\ref{star_formation_models}, with comparisons to the results of \citet{leroy_THINGS} included.  The total star formation activity of each galaxy is used to estimate the possible unseen H$_2$ content of each system in Section~\ref{H2_surf_dens_map}. The results of this paper are summarised in Section~\ref{summary}.

\section{NGC~2915 and NGC~1705}\label{n2915_n1705}
At a distance of $4.1\pm 0.3$~Mpc \citep{meurer2003} NGC~2915 is a nearby galaxy with the optical properties of a blue compact dwarf and the \hi characteristics of a late type spiral.  The optical disk has a $B$-band absolute magnitude of $M_B=-15.90$ and 25$th$ magnitude isophotal radii of 1.2~kpc and 1.9~kpc in the $B$- and $R$-bands respectively \citep{meurer1}, for the here adopted distance of 4.1~Mpc.   The stellar disk is contained within a very small region at the centre of a large \hi disk which extends out to $\gtrsim 22~B$-band scale lengths \citep{meurer2,elson_2010a_temp}.  The \hi disk has well-defined spiral structure as well as a bar-like feature co-located with the stellar disk.   No significant star formation is observed in the outer \hi disk, only a few faint \hii regions have been detected from deep H$\alpha$ imaging \citep{meurer_1999,werk_2010}.

Both \citet{meurer2} and \citet{elson_2010a_temp} have used the extended \hi disk of NGC~2915 to trace its gravitational potential out to large galactocentric radii.  Both investigations find the galaxy to be extremely dark-matter-dominated, with a total mass-to-blue-light ratio as high as $M_{tot}/L_B\sim 140$~\msun/$L_{B,\odot}$ \citep{elson_2010a_temp}.  The \hi velocity field of NGC~2915 was analysed in further detail by \citet{2915_flows} who find dynamical evidence for elliptical streaming and radial flow patterns within the \hi disk, as well as a possibly aspherical dark matter halo.  \citet{bureau_1999} showed that the \hi spiral structure can be caused by gravitational torques associated with a slow-rotating tri-axial dark matter halo.

NGC~1705, another blue compact dwarf, is well-known for hosting one of the most intense starbursts (relative to its mass) in the local universe.  \citet{tosi_et_al_2001} used the HST to resolve the brightest red giant stars in the galaxy, thereby estimating a distance of $5.1\pm0.6$~Mpc.  The system has a $B$-band absolute magnitude of $M_B=-15.6\pm0.2$ \citep{marlowe_1999}.   \citet{meurer_1705_1992} identified two stellar populations in NGC~1705 which they refer to as the high and low surface brightness populations.  The young stellar disk has a $B$-band 25$th$ magnitude isophotal radius of 1.2~kpc for the here adopted distance of 5.1~Mpc.  The galaxy's intense star formation activity is concentrated in its central high surface brightness stellar population, and is driven mainly by a powerful super star cluster, NGC~1705~-~1 \citep{sandage_1978,melnick_1985}, which contributes almost half of the total ultra-violet luminosity of the galaxy \citep{meurer_1705_2}.  \citet{Annibali_2003} found evidence for star formation activity that commenced at least 5~Gyr ago.  The authors also showed the star formation history of NGC~1705 to be very complex, confirming the existence of a recent burst of star formation between 10 and 15~Myr ago.  Consistent with the HST observations of NGC~1705~-~1, \citet{Annibali_2003} found the young and intermediate-aged stars to be concentrated near the very centre of the galaxy.  

Like NGC~2915, the stellar disk of NGC~1705 is embedded in a large, extended \hi disk.  The first \hi synthesis observations of the galaxy were carried out by \citet{meurer_1705_2} who provided dynamical evidence for a galactic blowout that is powered by the central super star cluster.  No distinct spiral structure is visible in the \hi distribution.  The general \hi kinematics of NGC~1705 are those of a rotating disk.  The dark matter properties of the galaxy were first modelled by \citet{meurer_1705_2} who showed the system to be dark-matter-dominated at nearly all radii.

\section{Data}\label{data}

Both galaxies have been observed at 21~cm, infrared, optical, and ultra-violet wavelengths.  This section presents and describes the multi-wavelength data sets utilised in this paper.

\subsection{\hi line observations}
\citet{meurer2} and \citet{meurer_1705_2} used the Australia Telescope Compact Array (ATCA) to carry out the first \hi synthesis observations of NGC~2915 and NGC~1705, respectively.  \citet{meurer2} used their data to produce the first mass models of a blue compact dwarf.  In this work we utilise new \hi synthesis observations from the ATCA with significantly improved sensitivity and spatial resolution.  

\begin{figure}
	\begin{centering}
	\includegraphics[angle=0,width=\columnwidth]{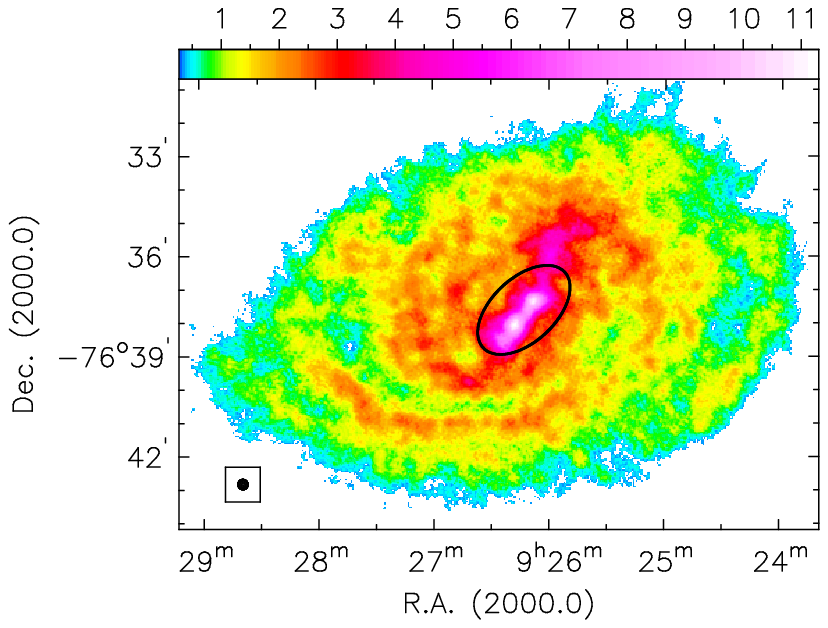}
	\includegraphics[angle=0,width=\columnwidth]{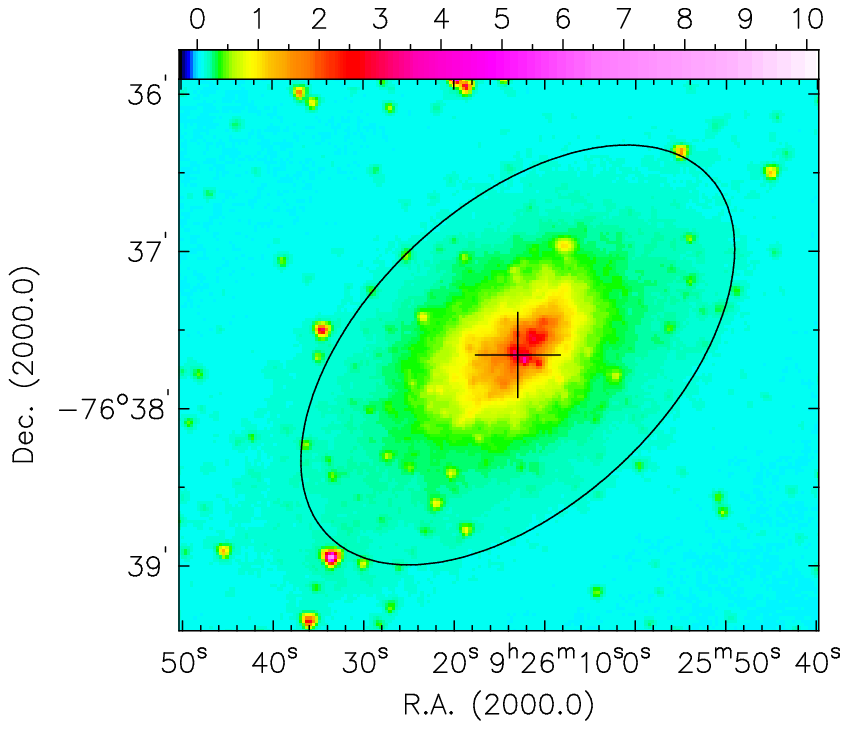}
	\caption{Top panel: NGC~2915 \hi surface density map produced using the data of \citet{elson_2010a_temp}.  The colour scale is described in units of \msun~pc$^{-2}$ by the colour bar.  The black ellipse represents the $R$-band $R_{25}$ radius of 1.9~kpc.  The hatched circle in the lower left corner of the panel represents the half power beam width of the synthesised beam.  Bottom panel: 3.6~\micron\ IRAC \emph{Spitzer} image of the old stellar disk of NGC~2915.  The black ellipse is the same as that in the panel above.  The cross marks the position of the photometric centre; $\alpha_{2000}$~=~09$^\mathrm{h}$~26$^\mathrm{m}$~12.611$^\mathrm{s}$, $\delta_{2000}$~=~$-76$\deg~37$'$~37.80$''$ as determined by \citet{elson_2010a_temp}.  The colour scale is described in units of MJy~ster$^{-1}$ by the colour bar.}
	\label{NGC2915_IR_HI}
	\end{centering}
\end{figure}

The \hi line data for NGC~2915 are those of \citet{elson_2010a_temp}.  These data have a spatial resolution of $17.0''\times 18.2''$, a factor of $\sim2$ improvement on the \citet{meurer2} observations.  The r.m.s. of the noise in a channel is $\sim 0.6$~mJy~beam$^{-1}$, the channel width is 3.2~km~s$^{-1}$.  The reader is referred to \citet{elson_2010a_temp} for a full description of the observations setups, the data reduction procedures, and a presentation of the various \hi data products including the channel maps.  An \hi surface density map of NGC~2915 produced using the data from \citet{elson_2010a_temp} is shown in the top panel of Fig.~\ref{NGC2915_IR_HI}.  For reference, a 3.6~\micron\ IRAC \emph{Spitzer} image of the evolved stellar disk of the galaxy is shown in the bottom panel.  The \hi observations are able to resolve the central region of the \hi disk into two over-densities.  Also visible is the clear spiral structure as well as a plume-like feature extending from the centre of the galaxy towards the north-west.  It is this \hi surface density map that is used to investigate the star formation laws in this paper.

\begin{figure}
	\begin{centering}
	\includegraphics[angle=-90,width=1\columnwidth]{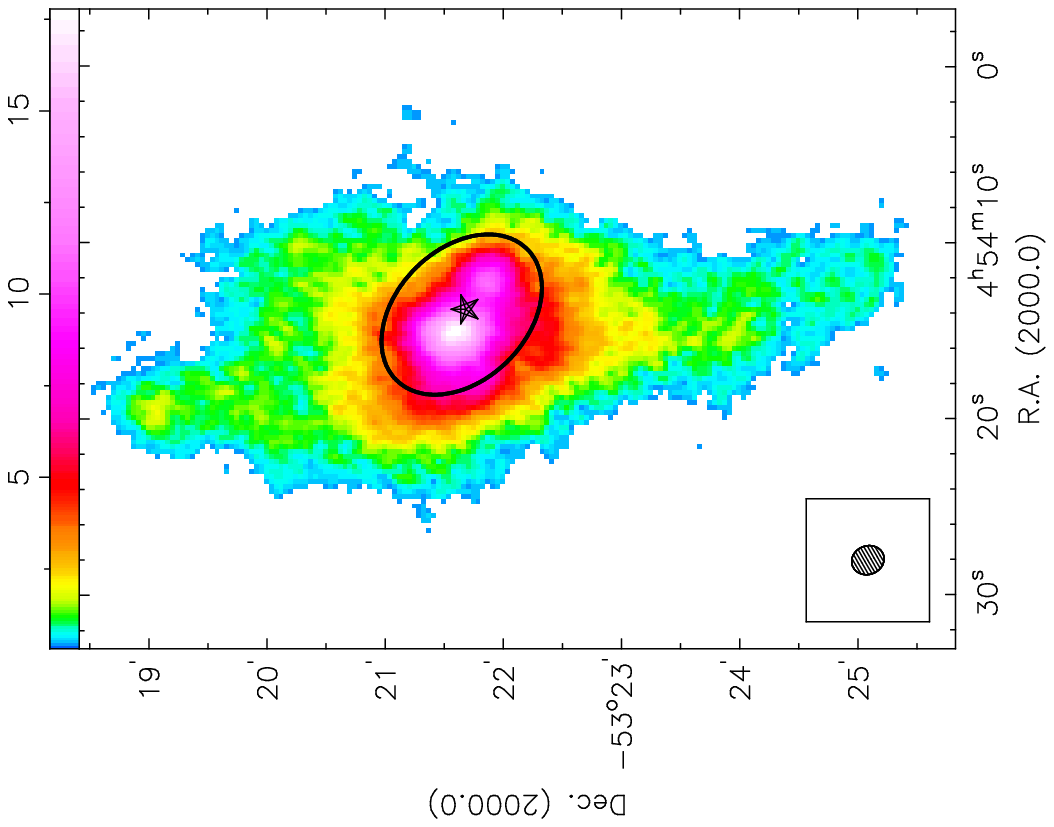}				
	\includegraphics[angle=0,width=1\columnwidth]{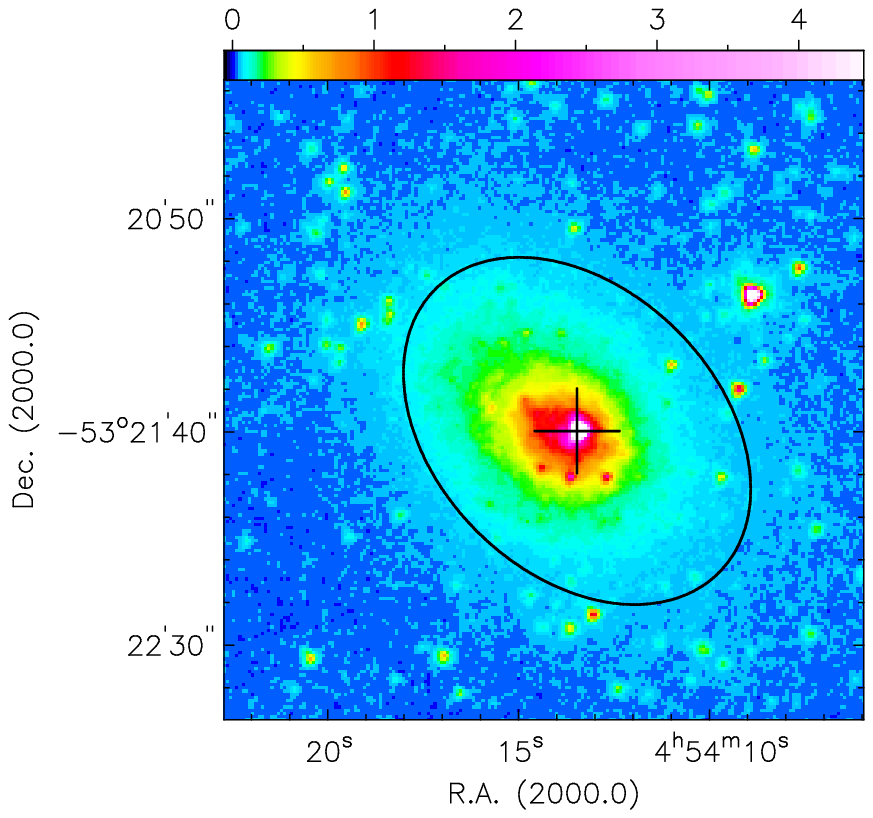}
	\caption{\small{Top panel: NGC~1705 \hi surface density map extracted from the new ATCA \hi synthesis data.  The colour scale is described in units of \msun~pc$^{-2}$ by the colour bar.  The black ellipse represents the $B$-band $R_{25}$ radius of 1.2~kpc.  The filled star marks the central position of the super star cluster, NGC~1705-1.  The hatched circle in the lower left corner of the panel represents the half power beam width of the synthesised beam.  Bottom panel: 3.6~\micron\ IRAC \emph{Spitzer} image of the old stellar disk of NGC~1705.  The black ellipse is the same as that in the panel above.  The cross marks position of the photometric centre; $\alpha_{2000}$~=~04$^\mathrm{h}$~54$^\mathrm{m}$~13.50$^\mathrm{s}$, $\delta_{2000}$~=~$-53$\deg~21$'$~39.80$''$ \citep{skrutskie_2006}.  The colour scale is described in units of MJy~ster$^{-1}$ by the colour bar.}}
	\label{NGC1705_IR_HI}
	\end{centering}
\end{figure}

The new NGC~1705 ATCA \hi synthesis observations used in this work will be discussed in detail in our upcoming publication (Elson~et~al 2012, in prep.).  For this work it suffices to say that approximately 86 hours worth of on source observations were obtained for the galaxy using baselines up to 6~km in length to produce an \hi data cube with a spatial resolution of $16.7''\times 14.5''$ and a channel width of 3.48~km~s$^{-1}$.  The r.m.s. flux of the noise in a line-free channel of the cube is $\sim 0.7$~mJy~beam$^{-1}$.   An \hi surface density map produced from these data is shown in the top panel of Fig.~\ref{NGC1705_IR_HI}.  A 3.6~\micron\ IRAC \emph{Spitzer} image of the stellar disk is again shown in the bottom panel of the figure for comparison.  The new \hi data clearly resolve the central \hi concentration of the galaxy into two over-densities with a combined mass of $\sim 3\times 10^7$~\msun.  These central \hi over-densities have their peaks separated by $\sim0.8$~kpc, and straddle the extremely luminous super star-cluster. 

\subsection{Ultra-violet and infrared imaging}
Both NGC~2915 and NGC~1705 have been observed with the GALEX satellite \citep{martin_GALEX_2005} as part of the GALEX Nearby Galaxy Survey \citep{GALEX}, and with the \emph{Spitzer} satellite as part of the \emph{Spitzer} Nearby Galaxies Survey \citep[SINGS,][]{SINGS}.  For each galaxy, the GALEX far ultra-violet (FUV) imaging and the \emph{Spitzer} 24~\micron\ imaging is used to trace the directly observable and dust-obscured SFRs, respectively.  These two tracers are combined to yield an estimate for the total SFR (Sec.~\ref{total_SFR}).

The GALEX FUV band, centred at 1450~\AA, covers the wavelength range 1350~-~1750~\AA.  The angular resolution for this band is $\sim~4.5''$.  The FUV data were photometrically calibrated and corrected for attenuation due to dust in the Galaxy.  The \citet{schlegel} dust extinction maps were used to estimate $E(B-V)\sim 0.^{\mathrm{m}}275$ and $E(B-V)\sim0.^{\mathrm{m}}008$ for NGC~2915 and NGC~1705, respectively.  These $B-V$ extinction measures were converted to FUV extinction magnitudes of $A_{FUV}=8.24\times E(B-V)=2.^{\mathrm{m}}26$ and $A_{FUV}=8.24\times E(B-V)=0.^{\mathrm{m}}06$ \citep{wyder_2007}.  Neither the FUV nor 24~\micron\ images of either galaxy suffer from significant star crowding effects.  No foreground star subtraction was performed.  

The MIPS instrument onboard the \emph{Spitzer} satellite has a field-of-view of $5.4'\times~5.4'$ in the 24~\micron\ band with a resolution of $\sim 6''$ and high signal-to-noise ratios.   \citet{bigiel_2008} allude to the fact that the MIPS point-spread-function at 24~\micron\ is severely non-Gaussian, yet presents no significant problems when the data are smoothed to resolutions of $\sim 20''$.  Again, no foreground star subtraction was performed for either galaxy. 

\begin{figure*}
	\begin{centering}
	\includegraphics[angle=-90,width=2\columnwidth]{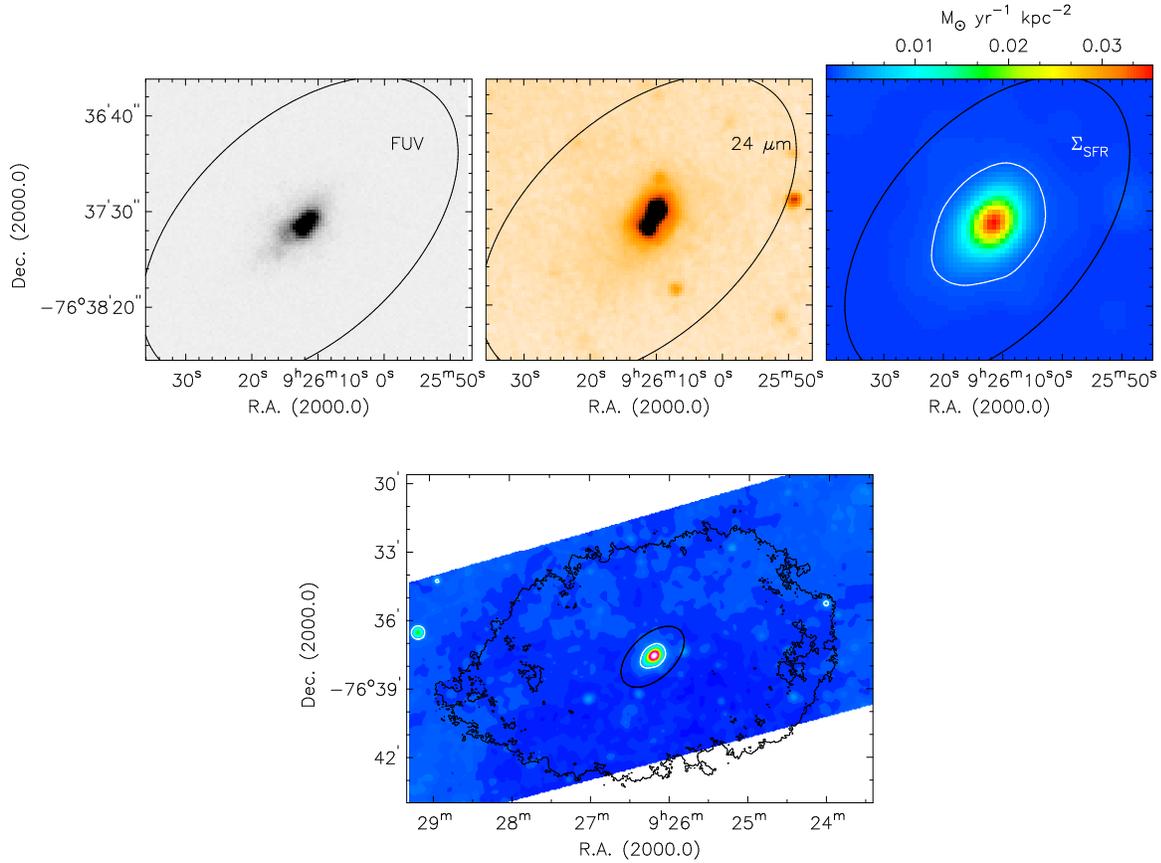}
	\caption{FUV, 24~\micron\ and total star formation rate surface density maps for NGC~2915.  Top left panel: GALEX FUV image of the stellar disk. Top middle panel: \emph{Spitzer} 24~\micron\ image of the stellar disk.  The FUV and 24~\micron\ images have been smoothed and re-gridded.  Top right panel: star formation rate surface density map for the stellar disk, constructed by combining the FUV and 24~\micron\ images according to Eqn.~\ref{SFR_eqn}.  Bottom panel: The full star formation rate surface density map.  The single black contour marks the edge of the \hi disk.  The white contour within the stellar disk is at a level of 0.0018~\msun~yr$^{-1}$~kpc$^{-2}$.}  
	\label{2915_FUV_MIPS_SFR}
	\end{centering}
\end{figure*}

\begin{figure*}
	\begin{centering}
	\includegraphics[angle=-90,width=2\columnwidth]{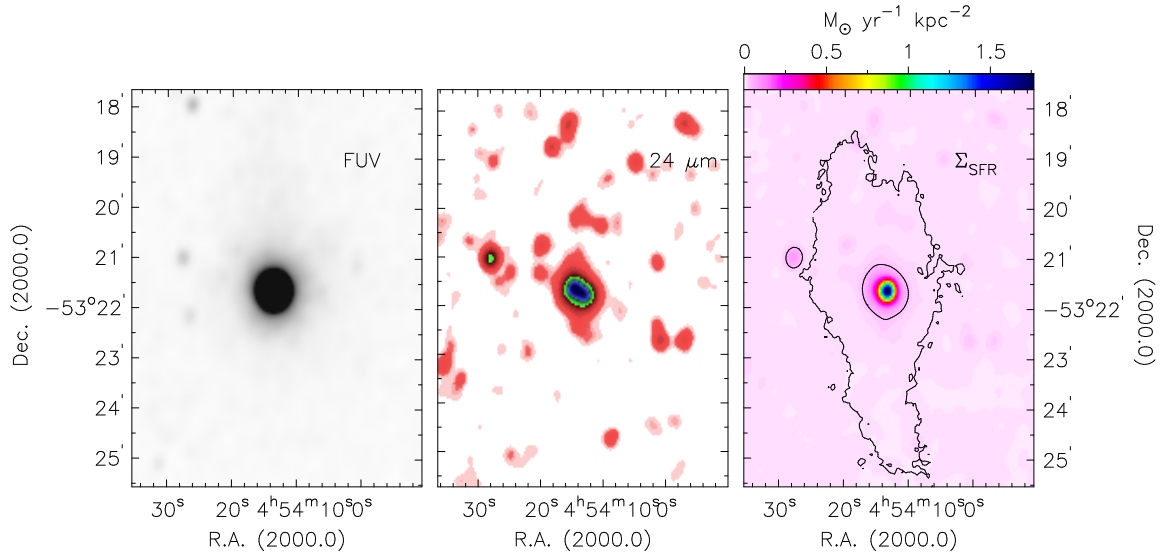}
	\caption{FUV, 24~\micron\ and total star formation rate surface density maps for NGC~1705. Left panel: GALEX FUV image of the stellar disk. Middle panel: \emph{Spitzer} 24~\micron\ image of the stellar disk.  The FUV and 24~\micron\ images have been smoothed and re-gridded.  Right panel: star formation rate surface density map constructed by combining the FUV and 24\micron\ images according to Eqn.~\ref{SFR_eqn}.  The outer black contour marks the edge of the \hi disk.  The inner black contour within the stellar disk is at a level of 0.1~\msun~yr$^{-1}$~kpc$^{-2}$.}  
	\label{1705_FUV_MIPS_SFR}
	\end{centering}
\end{figure*}

To facilitate a direct comparison of the FUV, 24~\micron\ and \hi data of each galaxy,  all images were placed on the same astrometric grid with the same spatial resolution.  The FUV and 24~\micron\ images of each galaxy were smoothed, using a Gaussian convolution function, to a resolution equal to the full-width-at-half maximum of the synthesised beam of the corresponding \hi data.  Next, these images were re-gridded to the same pixel size and astrometric grid as the \hi line data.  These smoothed, re-gridded FUV and 24~\micron\ images of NGC~2915 and NGC~1705 are shown in Figs.~\ref{2915_FUV_MIPS_SFR} and \ref{1705_FUV_MIPS_SFR}, respectively.

\section{Derived data}\label{2915_FUV_IR_SFR_maps_sec}
The aim of this work is to understand which properties of the neutral ISM regulate the star formation activity within each galaxy.  In order to do so we require quantitative measures of the star formation activity as well as parameterisations of the ISM kinematics.  This section describes our preferred method of deriving total star formation rate surface density estimates for each galaxy, and also the rotation curve and ISM velocity dispersion parameterisations required for the star formation models.

\subsection{The total star formation rate}\label{total_SFR}
The prescription of \citet{leroy_THINGS} is used to linearly combine the FUV and  24~\micron\ imaging of each galaxy to yield an estimate of the total SFR surface density, \sfrsd, in units of \msun~yr$^{-1}$~kpc$^{-2}$:
\begin{align}
{\Sigma_{SFR}\over M_{\odot}~\mathrm{yr^{-1}~kpc^{-2}}}&={\Sigma_{24~\mu m}\over M_{\odot}~\mathrm{yr^{-1}~kpc^{-2}}}\nonumber\\
&+{\Sigma_{FUV}\over M_{\odot}~\mathrm{yr^{-1}~kpc^{-2}}}
\\&= 3.2\times10^{-3}{I_{24}\over \mathrm{MJy~ster^{-1}}}\nonumber \\
&+8.1\times10^{-2}{I_{FUV}\over \mathrm{MJy~ster^{-1}}},
\label{SFR_eqn}
\end{align}
where $I_{24}$ and $I_{FUV}$ are the 24~\micron\ and FUV surface brightnesses, respectively.  Very importantly, \citet{leroy_THINGS} show that this choice of coefficients yields \sfrsd\ values that are consistent with the \Ha\--24~\micron\ calibration of \citet{calzetti_2007} and that when the 24~\micron\ emission is ignored the FUV \sfrsd\ estimates reduce to those of \citet{salim_2007}.  The calibration of \sfrsd\ uses the IMF from \citet{calzetti_2007} which is a Kroupa-type two-component IMF that extends to 120~\msun.  Some authors \citep[e.g.][]{meurer_IMF_2009} argue that the IMF is not universal, and that the slope of the upper end varies systematically.  A non-universal IMF implies that the SFR measured in a galaxy is highly sensitive to the tracer used in the measurement.  However, in this work we assume a standard two-component Kroupa IMF.  The \sfrsd\ maps for NGC~2915 and NGC~1705 are shown in Fig.~\ref{2915_FUV_MIPS_SFR} and Fig.~\ref{1705_FUV_MIPS_SFR}, respectively.  

\subsection{Rotation curve parameterisations}\label{vrot_params}
\begin{figure}
	\begin{centering}
	\includegraphics[angle=0,width=1\columnwidth]{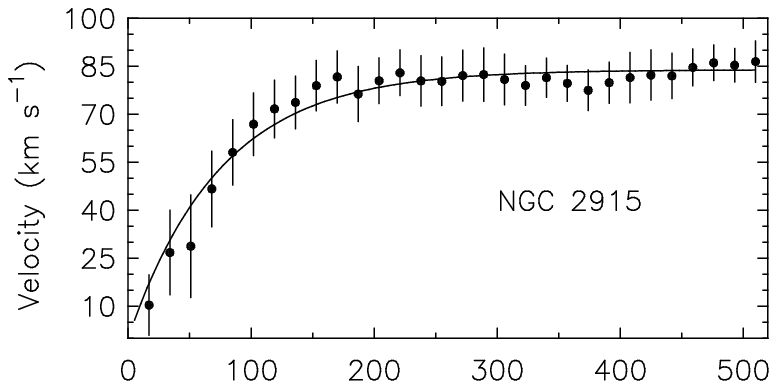}
	\includegraphics[angle=0,width=1\columnwidth]{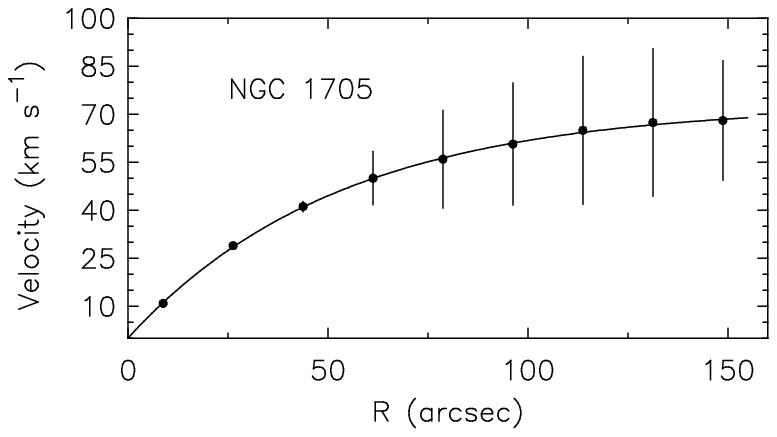}
	\caption{Observed rotation curves of NGC~2915 and NGC~1705 (black-filled circles) and their parameterisations (solid black curves).  Error bars in the top panel represent the r.m.s. spreads in the velocities within thin annuli of the fitted tilted ring model.  Error bars in the bottom panel represent the absolute difference in rotation velocity derived separately for each side of the galaxy.}
	\label{vrot_parameterisations}
	\end{centering}
\end{figure}

The star formation models investigated in this work require a measure of the galaxy's rotation curve as input.  To this end the rotation curves are parameterised using the function
\begin{equation}
 V(r)=V_{flat}[1-\exp(-r/l_{flat})],
\label{vflat}
\end{equation}
where $ V_{flat}$ and $ l_{flat}$ approximate the asymptotic velocity of the outer rotation curve and the length scale over which it approaches this constant velocity, respectively.   This parameterisation is easily analytically differentiable.  For NGC~2915, the rotation curve derived by \citet{elson_2010a_temp} is used.  This rotation curve was derived by fitting a tilted ring model to the \hi velocity field.  The best-fitting parameters are $V_{flat}=83.9$~\kms\ and $ l_{flat}=74.8''$ (1.5~kpc).  The rotation curve for NGC~1705 was generated by parameterising the \hi line profiles of an integrated position-velocity slice extracted from the new \hi data cube (the full details of the derivation of this rotation curve will be presented in our forthcoming publication Elson~et~al; 2012, in preparation).  This rotation curve is very similar to that derived previously by \citet{meurer_1705_2}, and is best parameterised by $V_{flat}=72.8$~\kms\ and $ l_{flat}=52.9''$ (1.3~kpc).      The rotation curves together with their respective parameterisations are shown in Fig~\ref{vrot_parameterisations}. 

\subsection{\hi velocity dispersions}
The star formation models also require an estimate of the gas velocity dispersion.  For some of the models considered in this work; \citet{kennicutt_1983}, \citet{kennicutt_1989} and \citet{martin_kennicutt_2001} used a constant gas velocity dispersion of $ \sigma_{gas}=6$~\kms, without distinguishing between the various thermal phases of the ISM.  Several authors have investigated and emphasised the importance of a \emph{cold} phase of the ISM for the stability of disk galaxies.  As an example, \citet{deblok_walter_2006} studied the \hi line profiles of NGC~6822 to identify separate warm and cold phases with velocity dispersions of $\sim$~8.2~\kms\ and $\sim$~4.4~\kms, respectively.  They showed that the velocity dispersion of the cold component, when used with a Toomre~Q criterion, gives an optimal description of ongoing star formation in NGC 6822, superior to that using the more conventional dispersion value of 6~\kms\ or the 8.2~\kms\ value of the warmer component.  \citet{schaye_2004} showed that the drop in thermal velocity dispersion associated with the transition from the warm ($T\sim 10^4$~K) to the cold ($T\sim 10^2$~K) phase of the ISM triggers gravitational instability on a wide range of scales.


Throughout this work we assume a constant value of $\sigma_{gas}=5.0$~\kms\ for the velocity dispersion of the cold gas phase of the neutral ISM.  Adopting a radially-varying velocity dispersion based on the \hi second-order moments would be difficult in the cases of NGC~2915 and NGC~1705 due to the fact that near their centres their \hi kinematics are not virialised.

\section{Star formation models}\label{star_formation_models}
This section deals with the various star formation models generated for NGC~2915 and NGC~1705.  Each model is introduced and its method of implementation discussed.  The results are presented together with a comparison to those of \citet{leroy_THINGS} who carried out similar analyses for their sample of 23 nearby star-forming galaxies from THINGS. This is done so that the results can be interpreted in the context of other typical star-forming dwarf galaxies.

\subsection{Single-fluid Toomre Criterion}\label{single_toomre}
\subsubsection{Introduction}
\citet{safronov_1960} was the first to show that perturbations in a thin, rotating gaseous disk can become unstable to gravitational collapse due to the effects of the self-gravity of the disk.   Building on the work of \citet{safronov_1960}, \citet{Toomre_1964} showed that a stellar disk that is fairly smooth or uniform, and that is rotating in approximate equilibrium between its self-gravitational and centrifugal forces, cannot be entirely stable against the tendency to gravitationally collapse.  This is the case unless the random motions within the disk are sufficiently large.  Today the Toomre~Q criterion is most commonly used to quantify the gravitational growth of perturbations within a thin, rotating gaseous disk.  According to this criterion, the disk should be unstable to axisymmetric disturbances in regions where the Toomre parameter,
\begin{equation}
Q\equiv{\sigma_{gas}\kappa \over \pi G\Sigma_{gas}}
\label{Q1}
\end{equation}
is less than unity.  The self-gravity, pressure and kinematics of the gas disk are represented by $ \Sigma_{gas}$, $ \sigma_{gas}$ and $ \kappa$, respectively.  $G$ is the gravitational constant.   The epicyclic frequency, $\kappa$, is a measure of the Coriolis or centrifugal forces stemming from the rotation of the perturbations.  Following \citet{kennicutt_1989}, $\kappa$ is estimated as
\begin{equation}
 \kappa(R)=1.41{V\over R}\sqrt{1+{R\over V}{dV\over dR}},
 \label{kappa_eqn}
\end{equation}
where $ V$ is the rotation velocity at a radius $R$.  It is the combination of random motions and centrifugal accelerations from epicyclic eddies that supports the ISM against collapse due to self-gravity.  The Toomre criterion therefore describes the abilities of perturbations to rotate around their centre of gravity and thus their stability against gravitational collapse.

\citet{kennicutt_1989} used a sample of 15 spiral galaxies to observationally test the Toomre criterion on galactic length-scales.  More recently, \citet{martin_kennicutt_2001} applied the prescription of \citet{Toomre_1964} to a sample of 32 star-forming spiral galaxies and compared the radial distributions of $Q_{gas}$ to \Ha\ radial surface brightness profiles.   For a sub-sample of 26 galaxies with well-defined thresholds they found a median value of $\alpha_Q=0.69\pm 0.2$  for the ratio of the gas surface density to critical surface density defined by Eqn.~\ref{Q1}, $\Sigma_{crit}=\sigma_{gas}\kappa/\pi G$.  Using this ratio as an empirical calibration for the Toomre criterion, Eqn.~\ref{Q1} becomes
\begin{equation}
 Q_{gas}\equiv{\alpha_Q\sigma_{gas}\kappa \over \pi G\Sigma_{gas}}.
\label{Q2}
\end{equation}
This empirical version of the Toomre stability criterion is used throughout this work.  Gravitational instability of the gas is predicted by $Q_{gas}<1$.

\subsubsection{Methodology}
The Toomre \qgas\ parameter was calculated for every resolution element in the \hi surface density map of each galaxy.  In assuming a constant \hi velocity dispersion, we have emulated the approaches adopted by \citet[][]{kennicutt_1989,kennicutt_1998,martin_kennicutt_2001,leroy_THINGS}.  The epicyclic frequency, $\kappa$, was determined using Eqn.~\ref{kappa_eqn}.  The parameterisation of the observed rotation curve of each galaxy was used to analytically determine ${dV\over dR}$.  An azimuthally averaged approximation of the epicyclic frequency was thus used for each of the resolution elements at a particular galactocentric radius.  The \qgas\ maps shown in this section can therefore be thought of as the \hi map modulated by a radial function that is largely determined by the rotation curve.  More sophisticated maps can be produced by allowing the \hi velocity dispersion to vary radially, but in this work we assume a constant velocity dispersion.

\subsubsection{Results and discussion}\label{2915_1705_1toomre_results}
\begin{figure}
	\begin{centering}
	\includegraphics[angle=0,width=1\columnwidth]{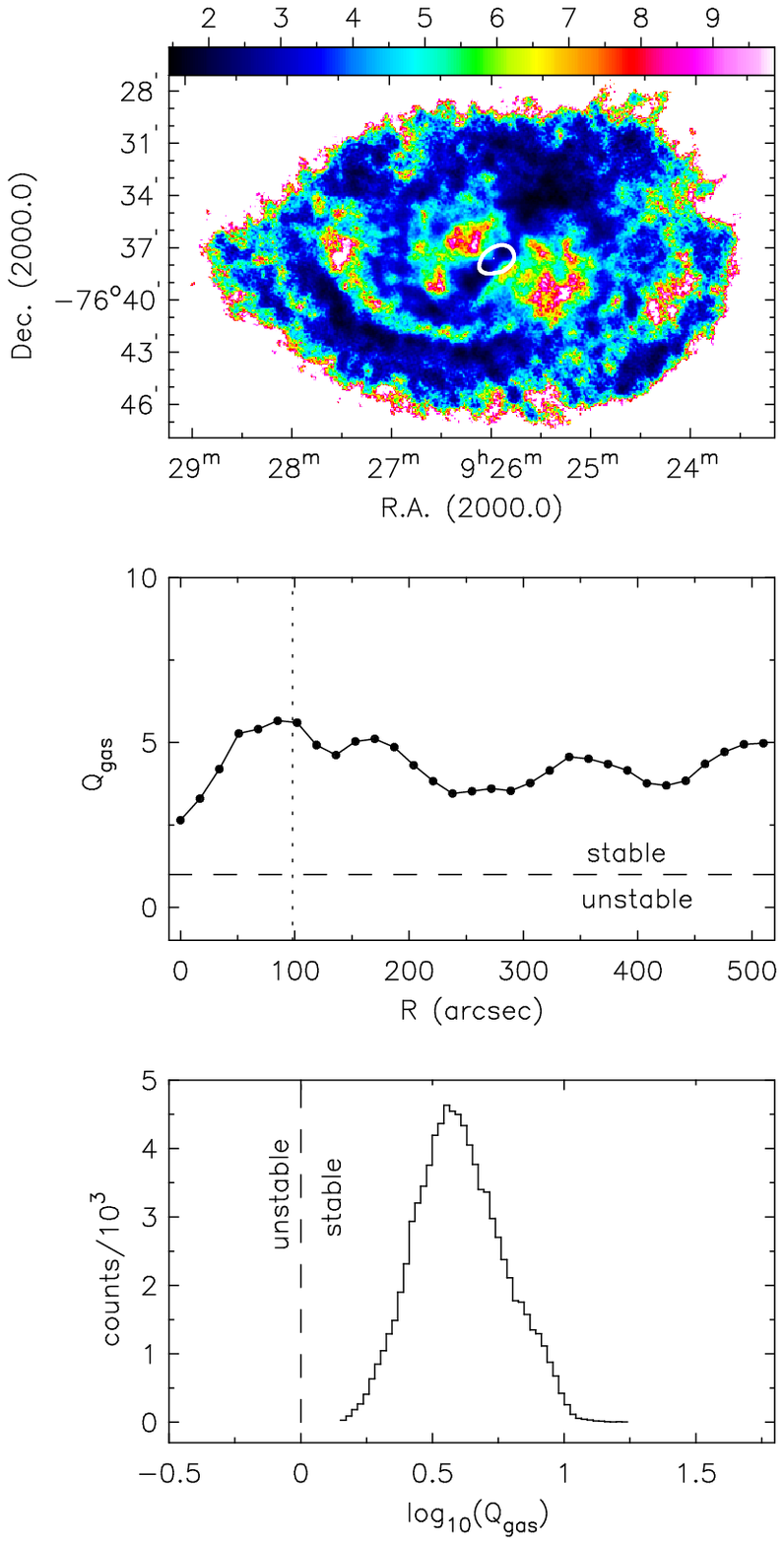}
	\caption{Top panel: \qgas\ map for the \hi disk of NGC~2915.  Middle panel: Radial profile of the \qgas\ map.  Bottom panel: Distribution of \qgas\ parameters.    The \qgas\ colour scale in the upper panel is descrbied by the colour bar.  The gas is expected to be unstable to gravitational collapse in regions where the \qgas$<1$.  The single white star formation rate surface density contour in the upper panel is at a level of \sfrsd~=~0.0018 \msun~yr$^{-1}$~kpc$^{-2}$.  The dotted vertical line shown in the middle panel represents the $R$-band $R_{25}$ radius at 1.9~kpc.}
	\label{2915_toomre_map}
	\end{centering}
\end{figure}

\begin{figure}
	\begin{centering}
	\includegraphics[angle=0,width=1\columnwidth]{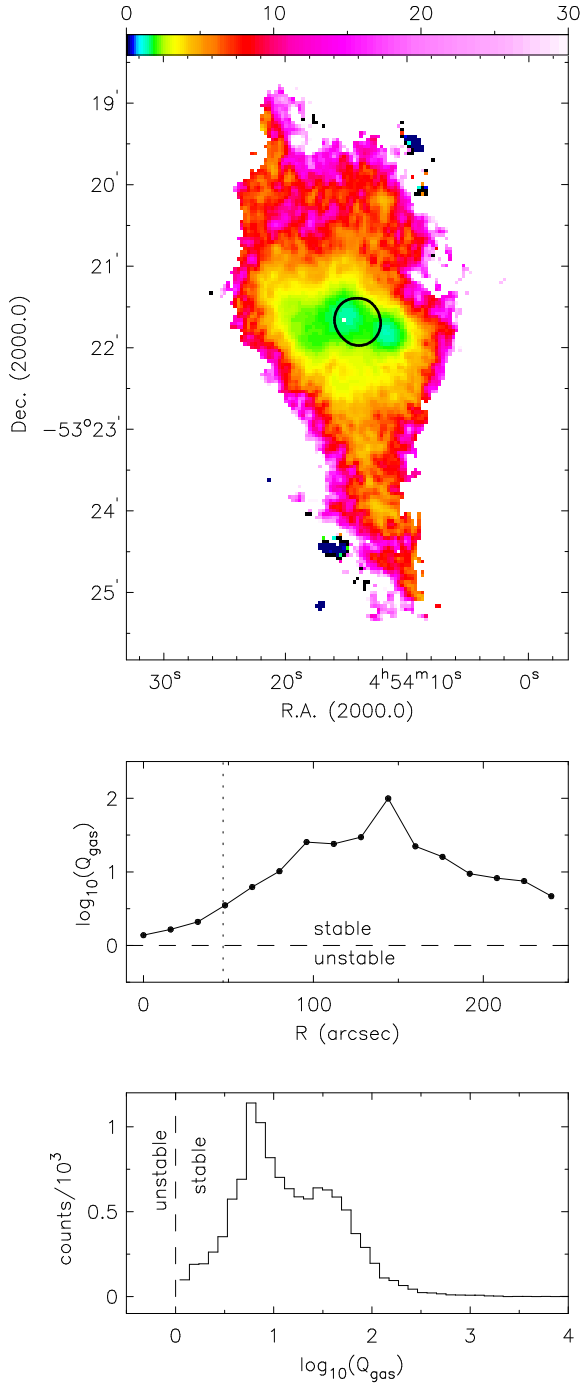}
	\caption{Top panel: \qgas\ map for the \hi disk of NGC~1705.  Middle panel: Radial profile of the \qgas\ map.  Bottom panel: Distribution of \qgas\ parameters.    The \qgas\ colour scale in the upper panel is described by the colour bar.  The gas is expected to be unstable to gravitational collapse in regions where the \qgas$<1$.  The single black star formation rate surface density contour in the upper panel is at a level of \sfrsd~=~0.1 \msun~yr$^{-1}$~kpc$^{-2}$.  The dotted vertical line shown in the middle panel represents the $B$-band $R_{25}$ radius at 1.2~kpc.}
	\label{1705_toomre_map}
	\end{centering}
\end{figure}

The results for NGC~2915 and NGC~1705 are shown in Figs.~\ref{2915_toomre_map} and \ref{1705_toomre_map}, respectively.  Most striking is the fact that, using the above-mentioned inputs, the Toomre criterion predicts the \hi disk of both galaxies to be sub-critical, that is $ Q_{gas}>1$ for every resolution element in the \hi total intensity map.  This result is inconsistent with the star-formation activity observed at the centre of each galaxy.  The lower panel of Fig.~\ref{2915_toomre_map} shows the logarithmic \qgas\ distribution for NGC~2915 which is  approximately Gaussian, peaking at $\log Q_{gas}\approx 0.55$, far-removed from \qgas~=~1 required for gravitational instability.  

Despite there being no $Q_{gas}\le1$ in the presence of clear star formation activity at the centre of NGC~2915,  the \qgas\ map does still exhibit the general structure that one might expect if it were to correctly predict the star formation activity.   The \qgas\ radial profile (Fig.~\ref{2915_toomre_map}, middle panel) shows the lowest \qgas\ values to occur at the centre of the galaxy, as well as a clear drop in the \qgas\ values at the edge of the stellar disk ($R= R_{25}$).  

The situation is similar for NGC~1705 whose logarithmic \qgas\ distribution is double-peaked with $Q_{gas}> 1$ throughout the \hi disk.  This galaxy's intense star formation activity is not well-described by the ~\qgas\ criterion.  The very large estimates of $Q_{gas}\sim 10^{1.5}$ for the outer \hi disk are consistent with the results of \citet{meurer_1705_2}.  As above, the lowest \qgas\ values again occur near the galaxy's central stellar disk despite the model predicting the \hi disk to be sub-critical.

\subsubsection{Comparison to other star-forming dwarfs}
How do the results of NGC~2915 and NGC~1705 compare to those of other star-forming dwarfs?  \citet{leroy_THINGS} found almost no area of the inner disks of their THINGS galaxies to be formally unstable to gravitational collapse.  Rather, they found that $Q_{gas}\sim4$ is typical for the region $R\lesssim 0.8R_{25}$ and that $Q_{gas}\gtrsim 10$ is common for larger radii.  In this sense, the results for NGC~2915 and NGC~1705 are consistent with the sample of \citet{leroy_THINGS}.  The authors find no clear evidence of a \qgas\ threshold that can unambiguously distinguish between star-forming regions of high and low efficiency.  \citet{leroy_THINGS} did, however, use $\sigma_{gas}=11$~\kms\ as their preferred measure of the kinetic support of the ISM.  They go on to point out that under various assumptions regarding the H$_2$ content and the thermal pressure of the galaxies, the median value of  \qgas\ in the outer disks of dwarfs agrees quite well with the threshold value of  \qgas\ determined by \citet{kennicutt_1989} and \citet{martin_kennicutt_2001}.  Our results for NGC~2915 and NGC~1705 as well as those of \citet{leroy_THINGS} are also similar to those of \citet{Hunter_1998} who, for their sample of dwarf galaxies, found \qgas\ to be systematically higher by a factor of $\sim2$ than the value of determined by \citet{kennicutt_1998}.

\subsection{Stars+Gas Two-Fluid Toomre Criterion}\label{stars_gas_sec}
\subsubsection{Introduction}
The Toomre criterion discussed in the previous section incorporates only the gravitational potential of the gaseous disk.  In addition to its gaseous potential, the stellar potential of a galaxy plays an important role in regulating the gravitational stability of the ISM by contributing to its self-gravity.  For this reason two-fluid instability models incorporating the gaseous \emph{and} stellar potentials are particularly relevant to systems in which the stellar and gas masses are comparable.  NGC~2915 and NGC~1705 have stellar masses of $\sim 3.2$ and $8.8\times 10^8$~\msun, respectively.  Each galaxy also contains at least 10$^8$~\msun\ of \hi.  This section therefore deals with two-fluid instability models that incorporate the stellar \emph{and} gaseous potentials of NGC~2915 and NGC~1705.

\citet{rafikov_2001} determined the instability condition for a thin, rotating disk composed of gaseous and stellar components to be
\begin{equation}
 {1\over Q_{gas,*}}\equiv {2\over Q_*}{q\over 1+q^2}+{2\over Q_{gas}}R_{\sigma}{q\over 1+q^2R_{\sigma}^2}>1.
\label{rafikov_eqn}
\end{equation}
In this equation $Q_*\equiv \kappa\sigma_{*,r}/ \pi G\Sigma_*$, $ q\equiv k\sigma_{*,r}/\kappa$, with $k$ being the wave number of the instability.  The \emph{radial} velocity dispersion of the stars in the plane of the stellar disk is represented by $\sigma_{*,r}$, and $R_{\sigma}\equiv \sigma_{gas}/\sigma_{*,r}$.

\subsubsection{Methodology}
\begin{figure}
	\begin{centering}
	\includegraphics[angle=0,width=\columnwidth]{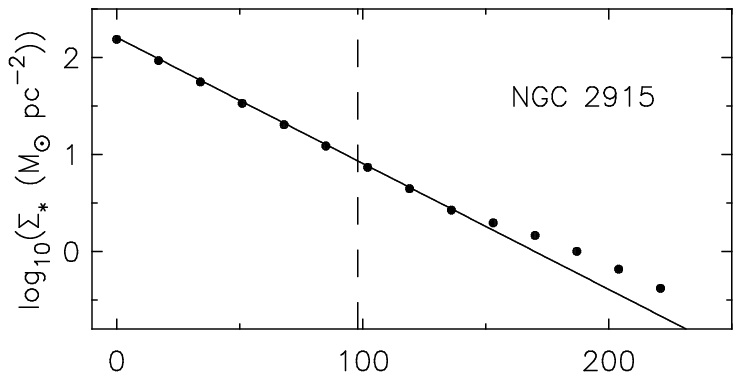}
	\includegraphics[angle=0,width=\columnwidth]{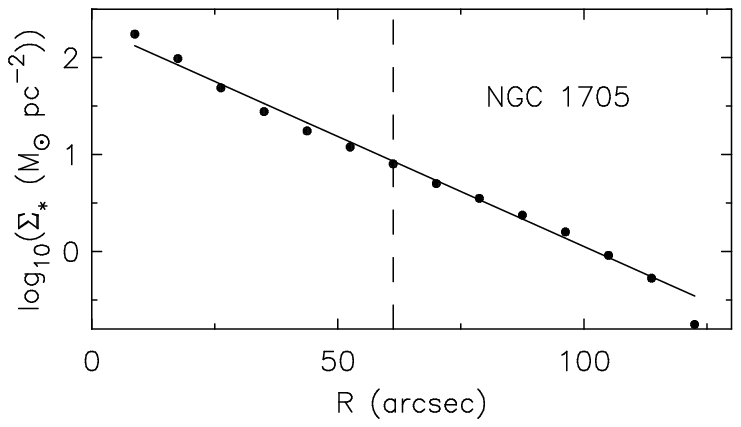}
	\caption{Stellar surface density radial profiles derived from the \emph{Spitzer} IRAC 3.6~\micron\ imaging of each galaxy (black filled circles).  All surface densities have been inclination corrected.  The parameterisation of each profile is shown as a solid black line.  The dashed vertical lines mark the $R_{25}$ isophotal radii of the galaxies.}
	\label{IR_profiles}
	\end{centering}
\end{figure}

The gravitational effects of the stellar mass component of each galaxy were determined using \emph{Spitzer} IRAC 3.6~\micron\ imaging.  Ellipses fitted to 3.6~\micron\ surface brightness isophotes of NGC~2915 suggest a constant inclination of $\sim 55^{\circ}$ and a position angle of $\sim 306^{\circ}$ for the old stellar population.  A radial profile (Fig.~\ref{IR_profiles}, top panel) was derived by azimuthally-averaging the stellar surface densities in thin annuli.  The parameterisation of this profile is
\begin{align}
\log_{10}\left({\Sigma_*\over M_{\odot}~pc^{-2}}\right)=-0.012{R\over \mathrm{arcsec}}+2.21.
\label{SSD1}
\end{align}
An identical procedure was carried out to generate a stellar surface density radial profile for NGC~1705 (Fig.~\ref{IR_profiles}, bottom panel).  A constant inclination and position angle of 45$^{\circ}$ was assumed.  The parameterisation of the resulting profile is
\begin{align}
\log_{10}\left({\Sigma_*\over M_{\odot}~pc^{-2}}\right)=-0.015{R\over \mathrm{arcsec}}+2.37.
\label{SSD2}
\end{align}
All stellar surface densities were inclination corrected so that Eqns.~\ref{SSD1} and \ref{SSD2} parameterise the face-on values.

The two-fluid stability criterion of \citet{rafikov_2001} requires an estimate of the radial component of the stellar velocity dispersion of each galaxy, $\sigma_{*,r}$.  The prescription of \citet{leroy_THINGS} was followed to determine $\sigma_{*,r}$ as a function of galactocentric radius.  \citet{leroy_THINGS} made four assumptions in order to determine the radial stellar velocity dispersion:
\begin{enumerate}
\item The scale height of an exponential stellar disk is constant with galactocentric radius.  This is a typical observation for edge-on galaxies \citep{kregel_2002}. 

\item A flattening ratio of $ l_*/h_*=7.3\pm 2.2$, where $l_*$ and $h_*$ are the disk scale length and scale height, respectively.  This is the average volume-corrected flattening of the disk light for the sample of 34 edge-on spiral galaxies of \citet{degrijs_1998} as determined by \citet{kregel_2002}.  From the IRAC 3.6~\micron\ imaging the stellar disk scale lengths of NGC~2915 and NGC~1705 were determined to be $ l_*\sim 0.6$~kpc and $ l_*\sim 0.14$~kpc, respectively, leading to $ h_*\sim 0.08$~kpc and $ h_*\sim 0.02$~kpc.
  
\item The stellar disk is isothermal in the z-direction.  Under the assumption of hydrostatic equilibrium, this leads to $ h_*=1/2\sqrt{\sigma_{*,z}^2/2\pi G\rho_*}$ \citep{vanderkruit_searle_1981}, where $ \rho_*$ is the mid-plane stellar volume density.  Using $ \Sigma_*=4\rho_* h_*$ and eliminating $ \rho_*$ then yields $\sigma_{*,z}=\sqrt{2\pi G\Sigma_* h_*}$ \citep{vanderkruit_1988}, leading to  
\begin{equation}
 \sigma_{*,z}=\sqrt{{2\pi Gl_*\Sigma_*\over 7.3}}.
\end{equation}

\item A fixed ratio of $ \sigma_{*,z}/\sigma_{*,r}=0.6$ to relate the vertical and radial stellar velocity dispersions.  This assumption is based on limited available evidence for late-type systems \citep{shapiro_2003}.
\end{enumerate}

\begin{figure}
	\begin{centering}
	\includegraphics[angle=0,width=1\columnwidth]{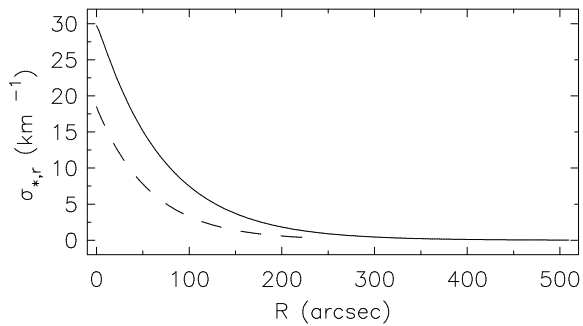}
	\caption{Radial profiles of the stellar velocity dispersion parallel to the plane of the stellar disk for NGC~2915 and NGC~1705 (solid and dashed curves, respectively), obtained by parameterising the stellar surface density radial profiles.}
	\label{sigmar_star}
	\end{centering}
\end{figure}

The parameterisation of the stellar surface density radial profile of each galaxy was used together with the $h_*$ approximation to yield $\sigma_{*,r}$ as a function of galactocentric radius.  Figure~\ref{sigmar_star} displays these $\sigma_{*,r}$ radial profiles.  At inner radii they are similar to the constant $ \sigma_{*,r}=25$~\kms\ estimate used by \citet{rafikov_2001} for the implementation and testing of his two-fluid stability criterion.  The radial stellar velocity dispersion monotonically approaches zero beyond the stellar disk.

\clearpage

So far, the prescriptions used to estimate $\Sigma_*$ and $\sigma_{*,r}$ have been presented.  These parameters together with the epicyclic frequency radial profile of each galaxy allow $Q_*$ in Eqn.~\ref{rafikov_eqn} to be determined.  The other parameters in Eqn.~\ref{rafikov_eqn} that need to be considered are $q$ and $R_{\sigma}$.  The $q$ parameter is a function of the wave number $k=2\pi/\lambda$, of the instability.  Specifying $k$ determines $q$.  According to  \citet{rosolowsky_blitz_2004}, 80$\%$ of the Milky Way's molecular gas is contained within giant molecular clouds ranging in size from $\sim$~2~pc to~$\sim$~100~pc.  Since we have no a priori knowledge of the linear size of a typical giant molecular cloud in NGC~2915 or NGC~1705, a set of perturbation wave numbers was selected that spans this length scale range.  Perturbation length scales of $\lambda$~=~0.005, 0.1, 0.2, 0.3, 0.5, 0.8~kpc were considered, each resulting in a separate instability map.  These perturbation length scales correspond to wave numbers of $k$~=~40, 20, 10, 6, 4, and 2.5 times $\pi$.  In the context of the instability maps, the length scales we consider range in linear extent from that of a single resolution element to that of the entire gaseous disk.  Finally, $R_{\sigma}=\sigma_{gas}/\sigma_{*,r}$ was specified by assuming $\sigma_{gas}=~5.0$~\kms\ for all two-fluid instability maps and using the $\sigma_{*,r}$ profiles shown in Fig.~\ref{sigmar_star}.

\subsubsection{Results and discussion}
\begin{figure*}
	\begin{centering}
	\includegraphics[angle=0,width=2\columnwidth]{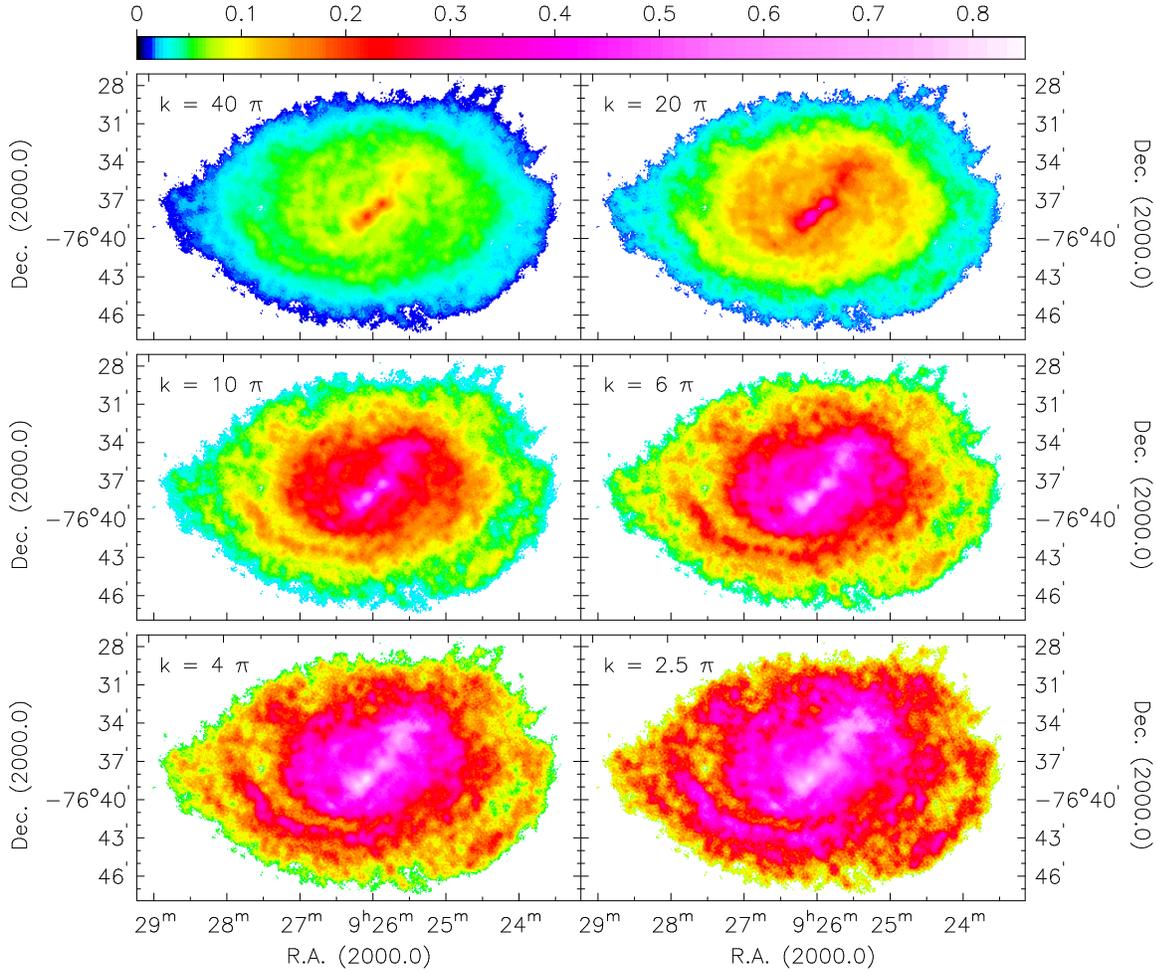}
	\caption{NGC~2915 stars+gas two-fluid instability maps for various perturbation length scales.  The common colour scale is specified by the colour bar at the top of the figure, and represents the two-fluid instability parameter, $ 1/Q_{gas,*}$.  The perturbation wave number, $k=2\pi/\lambda$, for each instability map is shown in the top left corner of the panel.  Wave numbers of $k$~=~40, 20, 10, 6, 4, 2.5 times $\pi$ correspond to wave lengths of $\lambda$~=~0.005, 0.1, 0.2, 0.3, 0.5, 0.8~kpc, respectively.  The gaseous disk is expected to be unstable to large-scale gravitational collapse in regions where $ 1/Q_{gas,*}>1$.  For all of the above-presented maps, the instability criterion predicts a sub-critical gaseous disk for NGC~2915.}
	\label{2915_2fluid_maps}
	\end{centering}
\end{figure*}

\begin{figure*}
	\begin{centering}
	\includegraphics[angle=0,width=2\columnwidth]{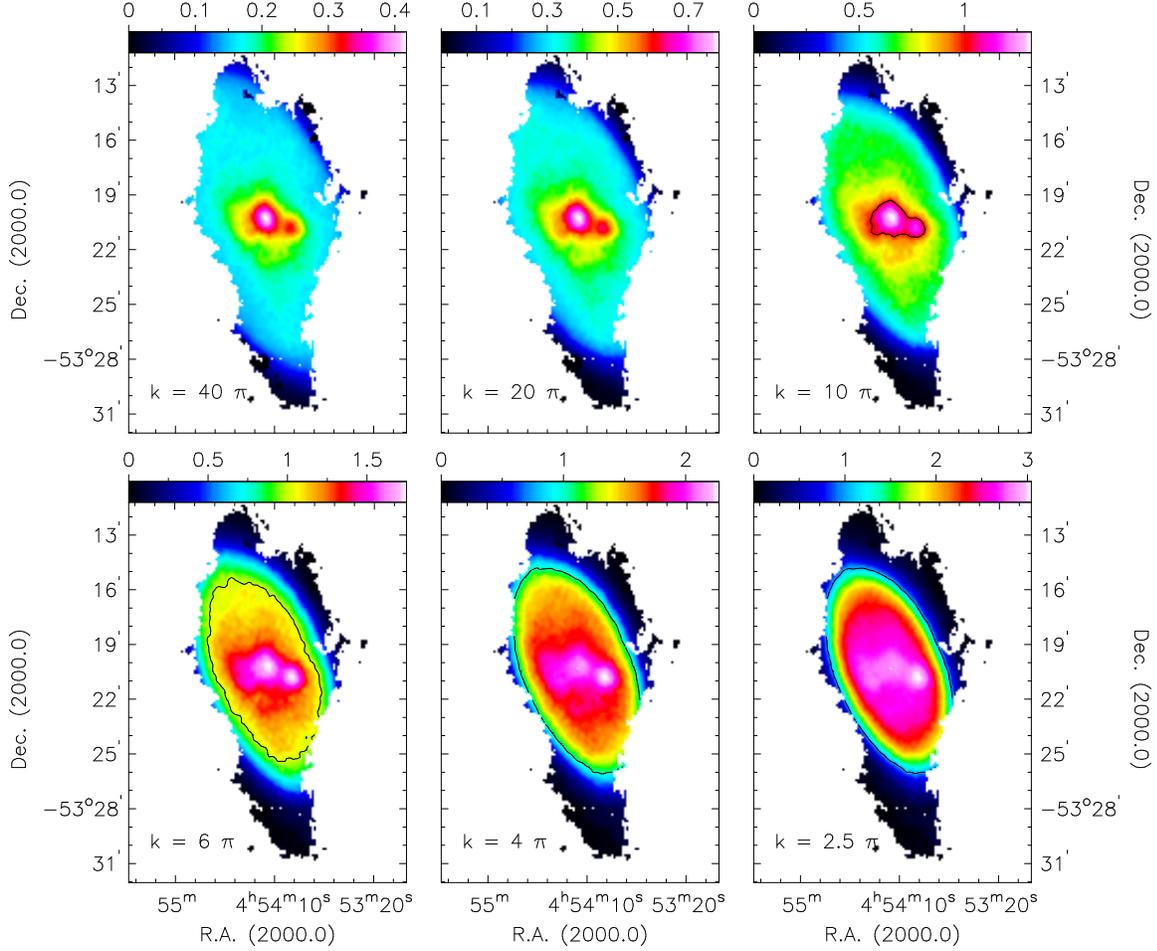}
	\caption{NGC~1705 stars+gas two-fluid instability maps for various perturbation length scales.  The intensity scale of each map is specified by the colour bar at the top of the panel, and represents the two-fluid instability parameter, $ 1/Q_{gas,*}$.  The perturbation wave number, $k=2\pi/\lambda$, for each instability map is shown in the top left corner of the panel.  Wave numbers of $k$~=~40, 20, 10, 6, 4, 2.5 times $\pi$ correspond to wave lengths of $\lambda$~=~0.005, 0.1, 0.2, 0.3, 0.5, 0.8~kpc, respectively.  The gaseous disk is expected to be unstable to large-scale gravitational collapse in regions where $ 1/Q_{gas,*}>1$.  The single black contour shown in some of the panels is at a level of $1/Q_{gas,*}=1$, enclosing the region of the galaxy in which gravitational instability is formally expected.}
	\label{1705_2fluid_maps}
	\end{centering}
\end{figure*}

\begin{figure}
	\begin{centering}
	\includegraphics[angle=0,width=1\columnwidth]{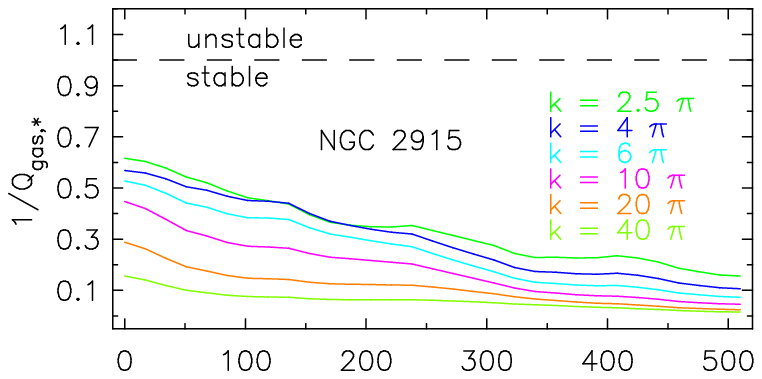}
	\includegraphics[angle=0,width=1\columnwidth]{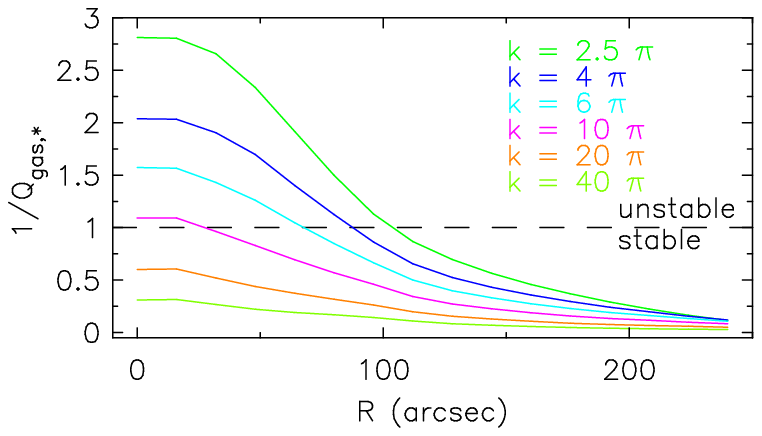}
	\caption{Radial profiles of the NGC~2915 and NGC~1705 stars+gas two-fluid instability maps shown in Figs.~\ref{2915_2fluid_maps} and \ref{1705_2fluid_maps}, respectively.  From top to bottom the profiles correspond to wave numbers of $k$~=~40, 20, 10, 6, 4, 2.5 times $\pi$.}
	\label{2fluid_profiles}
	\end{centering}
\end{figure}

\begin{figure}
	\begin{centering}
	\includegraphics[angle=0,width=1\columnwidth]{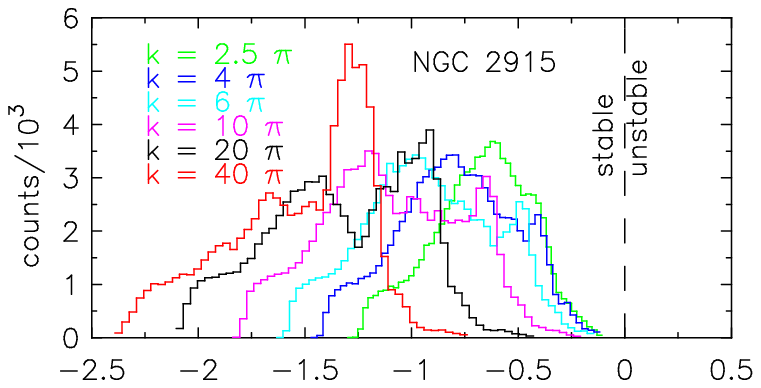}
	\includegraphics[angle=0,width=1\columnwidth]{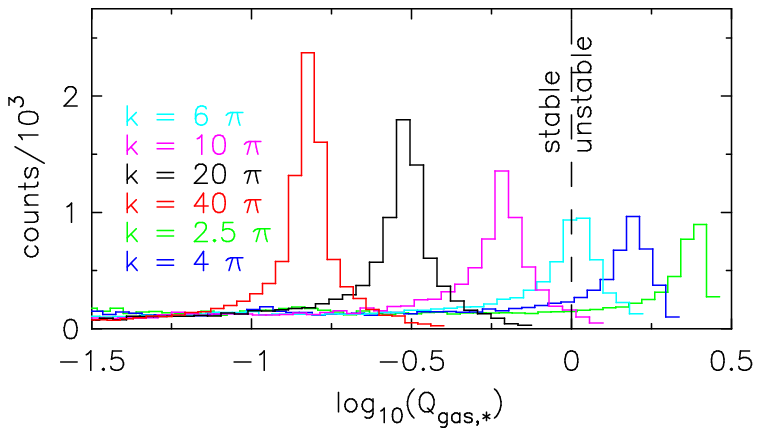}
	\caption{Distributions of the NGC~2915 and NGC~1705 stars+gas two-fluid instability maps shown in Figs.~\ref{2915_2fluid_maps} and \ref{1705_2fluid_maps}, respectively.}
	\label{2fluid_hists}
	\end{centering}
\end{figure}

The results of the $1/Q_{gas,*}$ two-fluid instability analyses for NGC~2915 and NGC~1705 are presented in Figs.~\ref{2915_2fluid_maps}~and~\ref{1705_2fluid_maps}.  The radial profiles of these maps are shown in Fig.~\ref{2fluid_profiles} while the distribution of $1/Q_{gas,*}$ is shown in Fig.~\ref{2fluid_hists}.  

The NGC~2915 results are considered first.  Even with the gravitational potential of the stellar disk included, the galaxy is sub-critical with $1/Q_{gas,*}~<~1$ throughout its \hi disk for all implemented wave numbers.  This two fluid instability criterion, like the single fluid instability criterion considered in the previous section, formally fails to predict the observed star formation activity at the centre of NGC~2915.  The extra self-gravity of the ISM induced by the stellar potential is still not enough to counteract the large epicyclic frequencies at inner radii.  Despite this formal failure for the central part of the galaxy, the instability maps again produce the sort of structure that one might expect for a late-type galaxy.  Beyond the radial extent of the stellar core, the two-fluid 1/\QGS\ maps resemble the single-fluid \qgas\ maps.  This result is expected since the stellar potential is negligible in the outer gaseous disk of the galaxy.  Including the gravitational influence of the stellar disk of NGC~2915 leaves the inner portion of its HI disk only marginally stable against gravitational collapse for the smallest wave numbers, corresponding to perturbation length-scales of $\sim0.5$~kpc.  This result is fairly consistent with that of \citet{leroy_THINGS}.  The authors found no clear evidence of a single-fluid Toomre~\qgas\ criterion that could unambiguously distinguish between high and low star formation efficiency regions.  However, when using Eqn.~\ref{rafikov_eqn} to include the effects of stellar gravity they found some regions became gravitationally unstable.  Including the stellar gravity led to large portions of the gaseous disks in their sample being only marginally stable against large-scale gravitational collapse, similar to our NGC~2915 result.  

The situation is only slightly better for the NGC~1705 instability maps which for some of the implemented wave numbers have $1/Q_{gas,*}>1$.  The $Q_{gas,*}=1$ level is marked with a single black contour in the instability maps shown in Fig.~\ref{1705_2fluid_maps}.  The best instability model is that for a wave number of $k=10\pi$, which corresponds to a perturbation length-scale of $\lambda=200$~pc.  This model predicts the very central regions of the \hi disk to be unstable.  In the context of a two-fluid Toomre criterion, the gravitational potential of the stellar disk seems to contribute enough self-gravity to the ISM to yield the central disk unstable.  Large perturbation length-scales of $\lambda\sim0.8$~kpc make the entire \hi disk of the galaxy unstable against gravitational collapse.  This demonstrates the inappropriateness of the perturbation size.  Perturbations with wave number $k\sim 10\pi$ yield only the inner disk unstable while the degree of instability is roughly constant over the outer \hi disk.  The 1/\QGS\ parameters for NGC~1705 generally span a small range of values.  As \citet{leroy_THINGS} point out, a small spread in 1/\QGS\ values is consistent with a self-regulating star formation scenario in which the gravitational potential associated with newly formed stars will seed further gravitational collapse of the ISM.  

Finally, it should be mentioned that several sources of uncertainty affect the results of these two-fluid instability analyses.  The main source is the lack of direct measurements of the stellar velocity dispersions.  The observed stellar surface density radial profile of each galaxy has been used to infer a stellar velocity dispersion profile.  This approach \emph{assumes} that $\sigma_*\propto\Sigma_*^{0.5}$ \citep{leroy_THINGS}.  This assumption may well affect the results of the star/gas two-fluid stability analyses.

\subsubsection{Comparison to other dwarfs}
A main finding of \citet{leroy_THINGS} was that considering the contribution of the stellar potential to the self-gravity of the ISM \emph{does not} render large, extended portions of the disk unstable.  The NGC~1705 results are well aligned with the \citet{leroy_THINGS} findings, with it having a significant fraction of its inner disk being formally unstable.  These results should be compared to the corresponding results for the simple Toomre~Q case in which only the gaseous potential was considered and for which no formal gravitational instability was predicted for either galaxy (i.e. $Q_{gas}>1$ throughout the disk).   Another main finding by \citet{leroy_THINGS} was that the 1/\QGS\ parameters exhibit a narrower range of values than the \qgas\ parameters.  This is clearly the case for NGC~1705 for which $1/Q_{gas,*}$ spans a range of $\sim0.3$~dex for all wave numbers.  \citet{leroy_THINGS} suggest that a narrow range of 1/\QGS\ values means that 1/\QGS\ offers little leverage to predict the locations of efficient star formation.  

\subsection{Dark matter+gas two-fluid Toomre criterion}\label{2toomre_criterionDM}
\subsubsection{Introduction}
The results of the previous section demonstrate the important role that the stellar gravitational potential of a galaxy can play in regulating the star formation activity.  This section attempts to answer the question of whether the dark matter (DM) can play a similar role.  

NGC~2915 and NGC~1705 are both DM-dominated galaxies.   NGC~2915 contains about $1.4\times10^{10}$~\msun\ of dark mass within $\sim~10$~kpc of its centre and has its DM strongly concentrated with a core density of $0.17$~\msunppc\ and a core radius of $r_c\sim 0.9$~kpc \citep{elson_2010a_temp}.  This is for the case in which the halo is modelled as a pseudo-isothermal sphere.  The situation is similar for NGC~1705 which contains $\sim3.1\times10^9$~\msun\ of DM within $\sim~6$~kpc of its centre, and which has a core density of $0.01$~\msunppc\ and a core radius of $1.2$~kpc (Elson~et~al.~2012, in prep.).   Since these galaxies contain dense concentrations of DM where the active star formation is observed, it is feasible that the portion of the DM halo of each galaxy that is co-located with its \hi disk plays a role in regulating the star formation activity.  

For NGC~2915, \citet{bureau_1999} proposed the self-gravitational effects of a large dark matter component co-located with the system's \hi disk as a possible explanation for its observed \hi spiral structure.  \citet{masset_bureau_2003} used hydrodynamical simulations to further explore this proposed mechanism, comparing the results to the observations using customised column density constraints.  They found that when the observed \hi density is scaled up by a factor of 10, the disk develops a spiral structure that closely resembles the observed one.  They suggest the disk of NGC~2915 to contain a large dark mass component that is closely linked to the observable neutral ISM.  They show further that the proposed scaling does not result in wide-spread star formation in the outer \hi disk, consistent with observations.  

\citet{Hunter_1998} state that disk dark matter ``effectively acts like stars in a two-fluid instability, giving extra self-gravity to small perturbations in the gas''.  In this section we therefore test an analogous version of the stars+gas instability criterion from Sec.~\ref{stars_gas_sec} by replacing the stellar quantities with corresponding DM quantities.  The two-fluid DM+gas instability criterion is
\begin{equation}
 {1\over Q_{gas,DM}}\equiv {2\over Q_{DM}}{q\over 1+q^2}+{2\over Q_{gas}}R_{\sigma}{q\over 1+q^2R_{\sigma}^2}>1,
\label{2fluid_DM_gas_eqn}
\end{equation}
where $Q_{DM}\equiv \kappa\sigma_{DM}/ \pi G\Sigma_{DM}$, $ q\equiv k\sigma_{DM}/\kappa$, with $k$ being the wave number of the instability.  The velocity dispersion of the DM is represented by $\sigma_{DM}$, and $R_{\sigma}\equiv \sigma_{gas}/\sigma_{DM}$.  This model, as compared to a simple Toomre~Q instability model with the \hi surface densities boosted to account for the disk dark matter, has the advantage that it incorporates a separate measure of the dark matter distribution and velocity dispersion.  The velocity dispersion of the dark matter is expected to be much larger than that of the \hi, meaning that equal masses of gaseous and dark matter will have different effects on the gravitational instability of the gas.

\subsubsection{Methodology}
\begin{figure}
	\begin{centering}
	\includegraphics[angle=0,width=1\columnwidth]{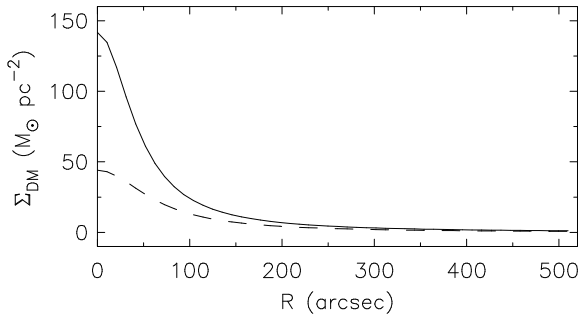}
	\caption{NGC~2915 and NGC~1705 dark matter surface density profiles (solid and dashed curves, respectively), as specified by Eqn.~\ref{dmsd_eqn}, used to construct the gas+DM two-fluid instability maps.}
	\label{DMprofile}
	\end{centering}
\end{figure}

$\Sigma_{DM}$ radial profiles for the galaxies are required by Eqn.~\ref{2fluid_DM_gas_eqn}.  Each  galaxy has its dynamics best explained by a pseudo-isothermal sphere parameterisation of its DM halo: 
\begin{equation}
\rho_{DM}(r)~=~{\rho_0\over 1+\left({r\over r_c}\right)^2},
\end{equation}
where $\rho_0$ is the DM core density and $r_c$ the core radius.  The DM surface density (in units of \msunpps) along a particular line of sight is modelled as the DM volume density at the corresponding galactocentric radius integrated over the thickness of the \hi disk:
\begin{eqnarray}
\Sigma_{DM}(r)&=&\int_{-h_{gas}}^{h_{gas}}\rho_{DM}(r)dh\\
&=&\int_{-h_{gas}}^{h_{gas}} {\rho_0\over 1+\left(r/r_c\right)^2}dh\\
&=&{2~h_{gas}~\rho_0\over 1+\left(r/r_c\right)^2}.
\label{dmsd_eqn}
\end{eqnarray}
\hi scale heights of $h_{gas}=0.41$ and 0.24~kpc are adopted for NGC~2915 and NGC~1705, respectively.  These estimates were generated using the prescription of \citet{IC2574}.  Adopting the $\rho_0$ and $r_c$ pseudo-isothermal sphere parameters mentioned in Sec.~\ref{2toomre_criterionDM}, the $\Sigma_{DM}$ radial profiles shown in Fig.~\ref{DMprofile} are obtained.

Also required by Eqn.~\ref{2fluid_DM_gas_eqn} is an estimate of the DM velocity dispersion.  In lieu of any such determination for either galaxy, we use the 3-dimensional velocity dispersion of an isotropic isothermal sphere:
\begin{equation}
\sigma_{DM}=\sqrt{{3\over 2}}V_{flat},
\end{equation}
where $V_{flat}$ is the asymptotic velocity of the outer rotation curve.  Adopting the best-fitting $V_{flat}$ parameters from Sec.~\ref{vrot_params}, $\sigma_{DM}$ esitmates of 102.7 and 89.2~\kms\ are obtained for NGC~2915 and NGC~1705, respectively.  Finally, whereas a range of wave numbers were tested in Sec.~\ref{stars_gas_sec}, here we test single wave numbers of $k=2.5\pi$ and $4\pi$ for NGC~2915 and NGC~1705, respectively.

\subsubsection{Results and discussion}
\begin{figure}
	\begin{centering}
	\includegraphics[angle=0,width=1\columnwidth]{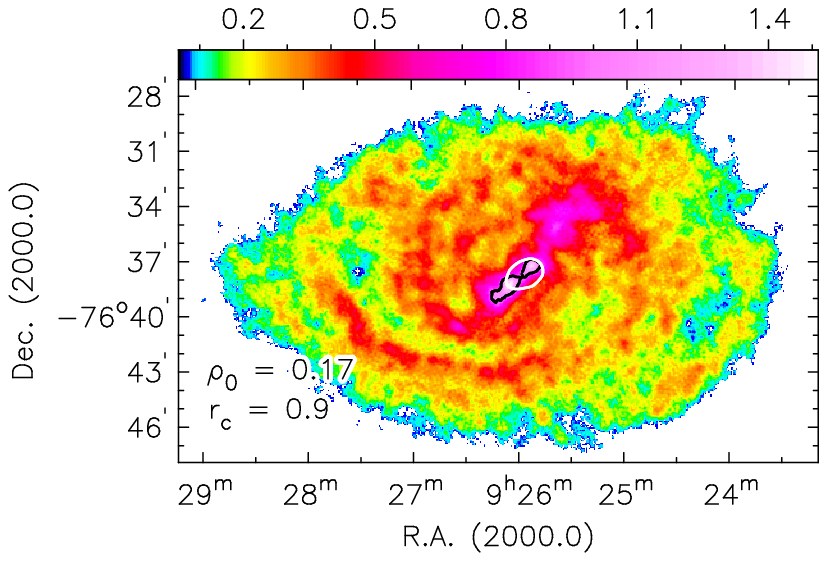}
	\caption{Gas+DM two-fluid instability map for NGC~2915.  The colour bar describes the $1/Q_{gas,DM}$ values.  The pseudo-isothermal parameters used to generate the map are specified in the lower left corner.  $\rho_0$ and $ r_c$ are in units of \msun~pc$^{-3}$ and kpc, respectively.  The edge of the stellar disk is delimited by the solid white \sfrsd~=~0.0018~\msun~pc$^{-2}$ contour while the gravitationally unstable portion of the gas disk is outlined by a solid black $ 1/Q_{gas,DM}=1$ contour.}
	\label{2915QGDM_maps}
	\end{centering}
\end{figure}

\begin{figure}
	\begin{centering}
	\includegraphics[angle=0,width=0.98\columnwidth]{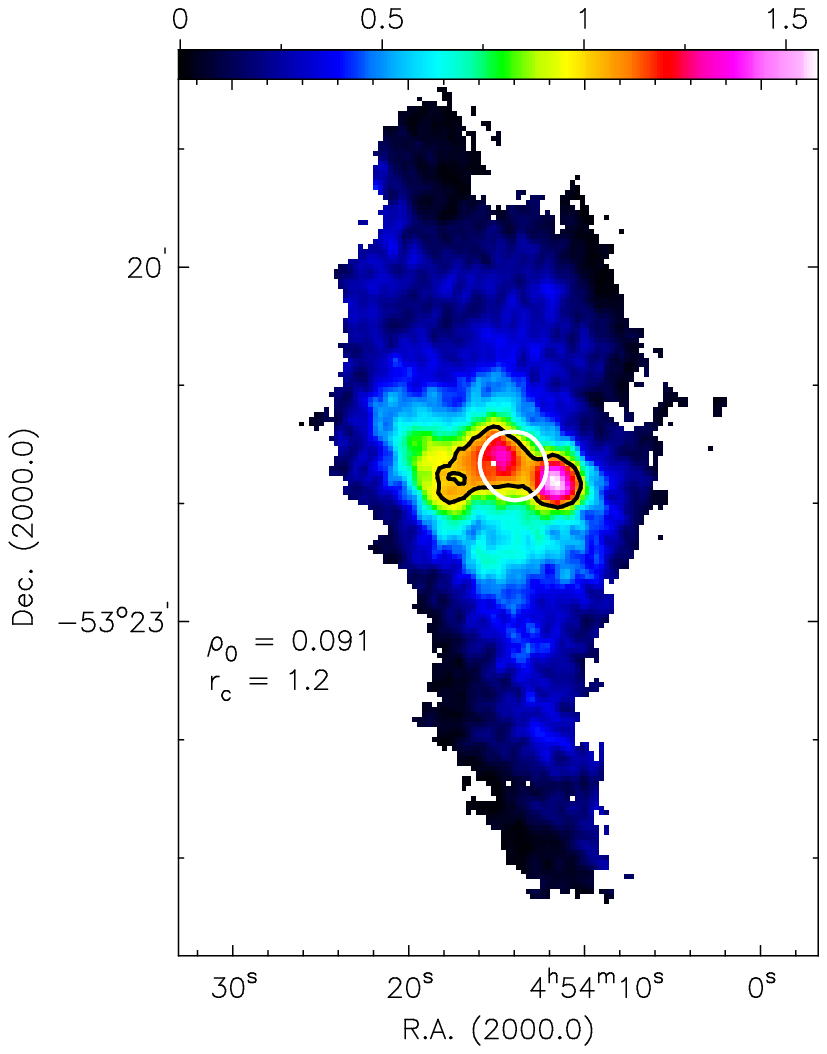}
	\caption{Gas+DM two-fluid instability map for NGC~1705.  The colour bar describes the $1/Q_{gas, DM}$ values.  The pseudo-isothermal parameters used to generate the map are specified in the lower left corner.  $\rho_0$ and $ r_c$ are in units of \msun~pc$^{-3}$ and kpc, respectively.  The edge of the stellar disk is delimited by the solid white \sfrsd~=~0.1~\msun~pc$^{-2}$ contour while the gravitationally unstable portion of the gas disk is outlined by a solid black $1/Q_{gas,DM}=1$ contour.}
	\label{1705QGDM_maps}
	\end{centering}
\end{figure}

The results for NGC~2915 and NGC~1705 are presented in Figs.~\ref{2915QGDM_maps} and \ref{1705QGDM_maps}, respectively.  The results suggest that part of the inner \hi disk of each galaxy can become unstable to gravitational collapse when additional self-gravity of the gas caused by the disk dark matter is incorporated.  Disk dark matter could therefore contribute towards regulating the star formation activity near the centres of NGC~2915 and NGC~1705.

Using each galaxy's pseudo-isothermal sphere parameterisation to estimate the amount of DM co-located with its \hi disk ensures that the inner \hi disk (where $\rho_{DM}(r)$ is high) contains more dark mass than the outer \hi disk (where $\rho_{DM}(r)$ is relatively lower).  The resulting distribution of $\Sigma_{DM}$ values is different to the one obtained by simply boosting each line-of-sight $\Sigma_{HI}$ measurement by a  constant factor.

In their review of cold gas accretion in galaxies, \citet{sancisi_review} allude to the fact that ``it is remarkable that there is such a pronounced spiral structure in the outer regions of spirals where dark matter dominates and even in dwarfs where the dark halo is believed to be predominant everywhere''.  The question, they therefore say, is ``whether these systems have light disks surrounded by massive dark halos or, rather, have heavy and dark disks''.  The modeling results from this section allow this question to be partly addressed for the cases of NGC~2915 and NGC~1705.  It has been shown that only a heavy disk scenario can account for the observed star formation in both NGC~2915 \emph{and} NGC~1705.  This result, together with the result of \citet{masset_bureau_2003} that a heavy disk allows for the natural formation of NGC~2915's observed \hi spiral structure suggests that certainly NGC~2915 may have a heavy \hi disk, and very likely NGC~1705 too, with significant amounts of dark mass distributed within each of them.

\subsection{Shear-based instability criterion}
A common feature of all the models presented so far is their characterisation of the global kinematics of a galaxy--they all incorporate the epicyclic frequency.  This section investigates a star formation model for which the kinematics are quantified not in terms of the Coriolis force (i.e. $\kappa$), but rather the rotational shear.

\subsubsection{Introduction}
\citet{Hunter_1998} suggest the Coriolis force incorporated in the Toomre~Q instability criterion to become less important in the presence of rotational shear. It is the time available for clouds to collapse in the presence of rotational shear, they argue, that regulates a galaxy's star formation activity.  \citet{Hunter_1998} use Oort's A constant,
\begin{equation}
A=0.5\left({V\over r}-{dV\over dr}\right),
\label{oort_A}
\end{equation}
to quantify the rotational shear of the gas.  Their so-called shear-based parameter for gravitational instability is then 
\begin{equation}
S_{gas}={\alpha_A\sigma_{gas}A \over \pi G\Sigma_{gas}}.
\label{sgas_eqn}
\end{equation}
Based upon the idea that a perturbation must grow by a factor of $\sim$~100 in the presence of shear for the instability to be significant,  \citet{Hunter_1998} estimate $\alpha_A\sim$~2.5.  This value of $A$ matches the contrast between the surface densities of neutral and molecular inter-stellar media in the presence of rotational shear.  It also allows the condition \qgas~$\lesssim 1$ to be met over the $V\propto R$ portion of the rotation curve. Regions within the galaxy that have $ S_{gas}<1$ should be forming stars while regions with $ S_{gas}>1$ should be stable against large-scale gravitational collapse.

\begin{figure}
	\begin{centering}
	\includegraphics[angle=0,width=1\columnwidth]{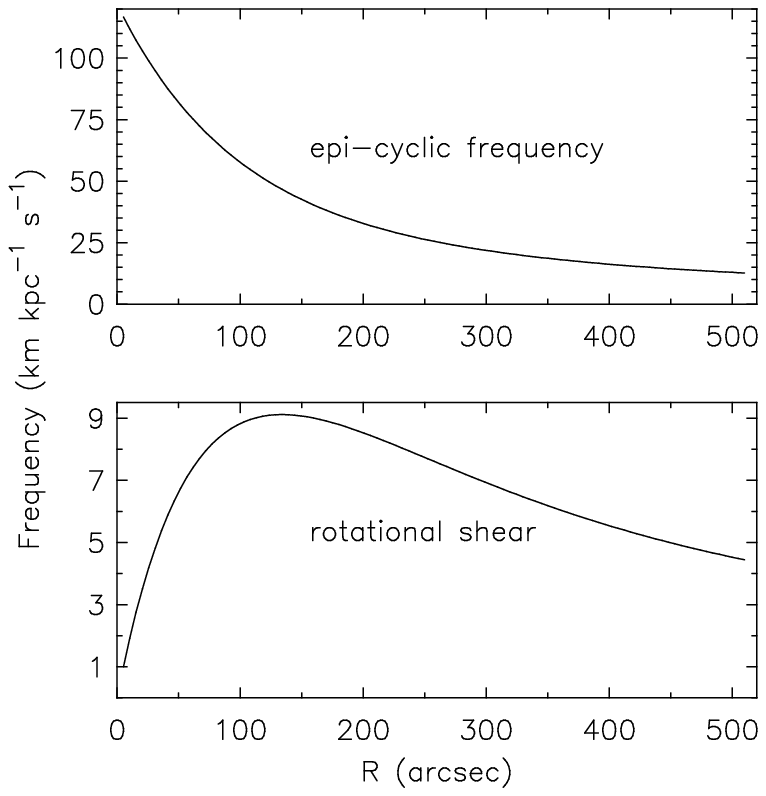}
	\caption{Epicyclic frequency and rotational shear based on the parameterisation of the rotation curve of NGC~2915 (Fig.~\ref{vrot_parameterisations}).}
	\label{VAK}
	\end{centering}
\end{figure}

\citet{Hunter_1998} point out that using $A$ to quantify the gas kinematics instead of $ \kappa$ makes very little difference for a flat rotation curve.  They estimate the two thresholds to be the same to within 12$\%$ for a flat rotation curve.  The difference between the two thresholds is most apparent for rising rotation curves where the rotational shear is low.  Under such circumstances $S_{gas}$ can be significantly lower than \qgas.  Figure~\ref{VAK} shows radial profiles of the epicyclic frequency and rotational shear for the parameterisation of the rotation curve of NGC~2915 (Fig.~\ref{vrot_parameterisations}).  Most noticeable is the fact that the rotational shear is a factor $\sim 5-30$ smaller in magnitude than the epicyclic frequency within $R\lesssim 150''$.  The situation for NGC~1705 is very similar, and it is therefore within the spatial extent of the stellar disks of the galaxies that a the shear-based characterisation of the kinematics is expected to be most influential when incorporated into a star formation model.

\subsubsection{Methodology}
\sgas\ maps were produced for NGC~2915 and NGC~1705 according to Eqn.~\ref{sgas_eqn}.  As for $\kappa$ in the Toomre~\qgas\ criterion, the shear parameter, $A$, for each line of sight was approximated as the azimuthally-averaged shear corresponding to the galactocentric radius of the resolution element.  The shear radial profile was determined for each galaxy using Eqn.~\ref{oort_A} together with the parameterisation of the galaxy's rotation curve.  

\subsubsection{Results and discussion}
\begin{figure}
	\begin{centering}
	\includegraphics[angle=0,width=1\columnwidth]{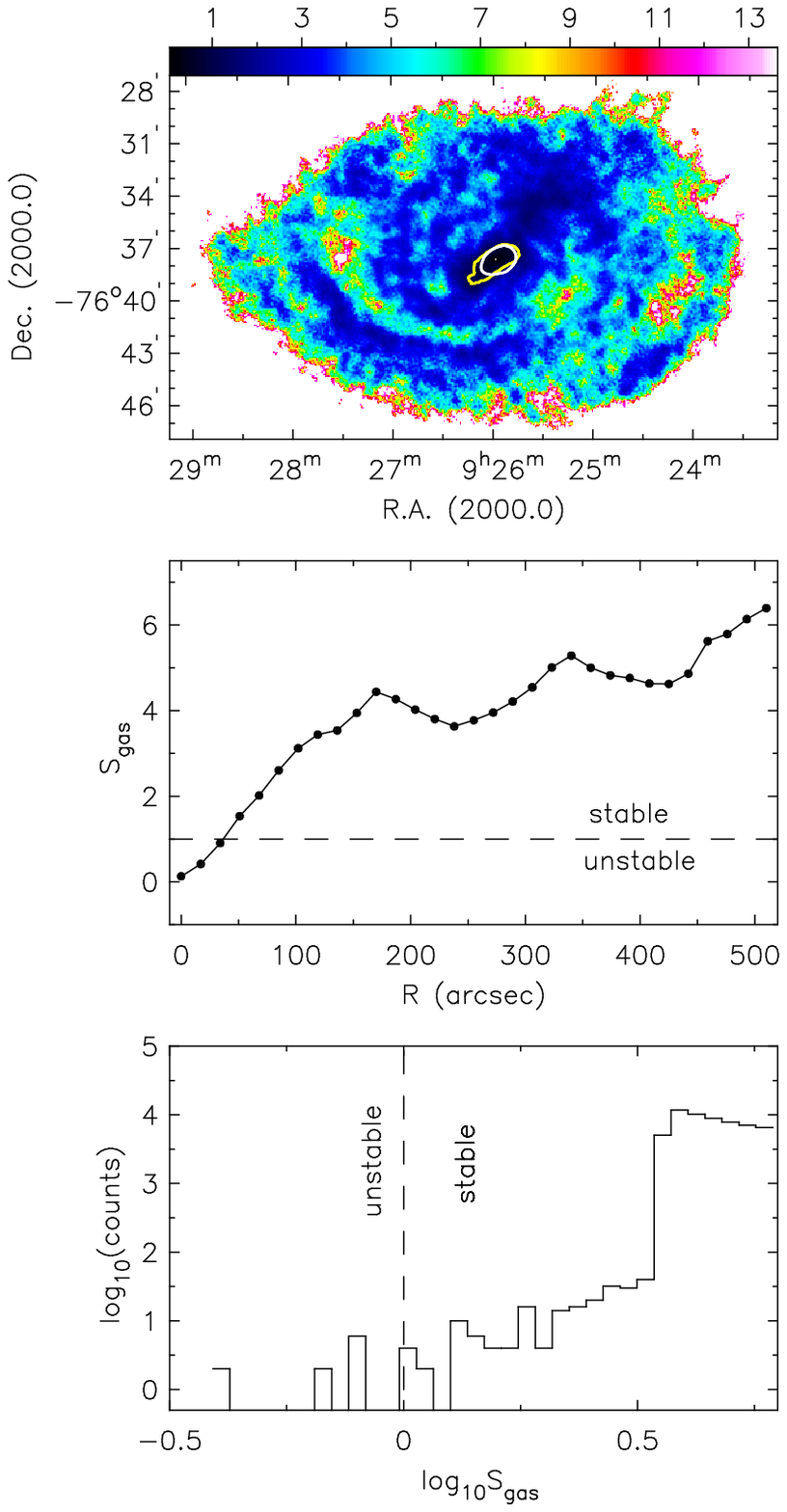}
	\caption{Top panel: \sgas\ instability map for the \hi disk of NGC~2915.  Middle panel: Radial profile of the \sgas\ instability map.  Bottom panel: Distribution of $\log_{10}(S_{gas})$ values.  The colour bar above the upper panel describes the \sgas\ values.  The single yellow contour in the map is at a level of $S_{gas}=1$, and encloses the unstable region of the \hi disk.  The single white contour in the upper panel, at a level of \sfrsd~=~0.0018 \msun~yr$^{-1}$~kpc$^{-2}$, approximates the edge of the stellar disk of NGC~2915.}
	\label{2915shear_map}
	\end{centering}
\end{figure}

\begin{figure}
	\begin{centering}
	\includegraphics[angle=0,width=1\columnwidth]{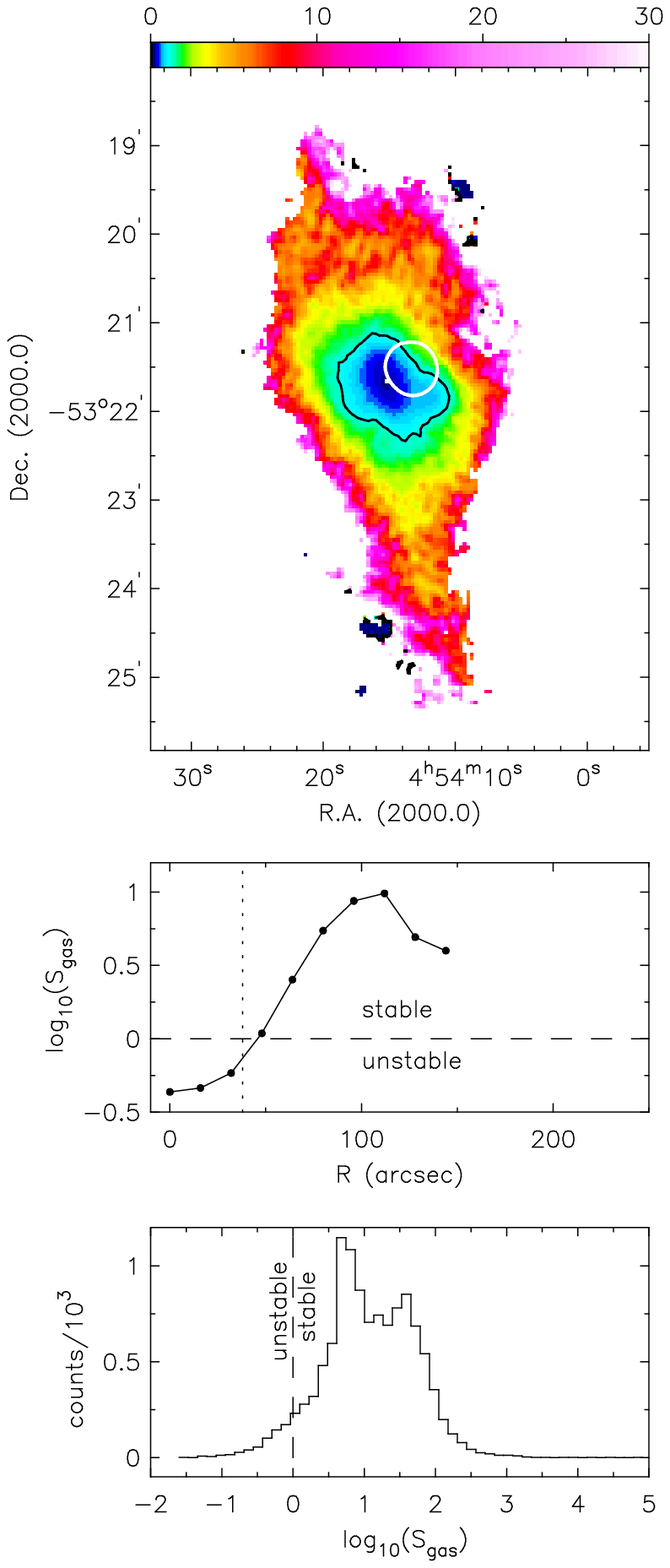}
	\caption{Top row: \sgas\ instability map for the \hi disk of NGC~1705.  Middle row: Radial profile of the $\log_{10}(S_{gas})$ values.  Bottom row: Distribution of $\log_{10}(S_{gas})$ values.  The colour bar above the upper panel describes the \sgas\ values.  The single black contour in the map is at a level of $S_{gas}=1$, and encloses the unstable region of the \hi disk.  The single white contour in the upper panel, at a level of \sfrsd~=~0.1 \msun~yr$^{-1}$~kpc$^{-2}$, approximates the edge of the star-forming core of NGC~1705.}
	\label{1705shear_map}
	\end{centering}
\end{figure}

The \sgas\ maps for NGC~2915 and NGC~1705 are shown in Figs.~\ref{2915shear_map} and \ref{1705shear_map}, respectively.  The \sgas\ criterion \emph{correctly} describes the star formation activity at the centre of each galaxy's gaseous disk.  The $S_{gas}=1$ contours of the shear maps are able to accurately trace the edges of the stellar disks of the galaxies.  By simply using a shear-based characterisation of the global dynamics, the \sgas\ criterion can correctly locate the unstable portions of the disks without the need for extra stellar or dark matter self-gravity.  There appears to be no tight correlation between \sgas\ and the observed star formation activity.  Within the central star-forming region of each galaxy, the \sgas\ parameters vary significantly (e.g. a factor of $\sim10$ for the case of NGC~1705's stellar disk).

\subsubsection{Comparison to other dwarfs}
The NGC~2915 and NGC~1705 findings are consistent with the results of \citet{leroy_THINGS} who found the inner disks of their galaxies to be more nearly super-critical in the context of a shear-based star formation threshold than in the context of a \qgas\ criterion.  Consistent with our results for NGC~2915 and NGC~1705 is the fact that \citet{leroy_THINGS} find a simple shear criterion to perform better than the star+gas two-fluid criterion for their sample of THINGS galaxies.  \citet{leroy_THINGS} found that, according to this shear-based instability criterion, there are regions of high star formation activity within the disk that are formally unstable against the tendency to gravitationally collapse, instead of being formally stable.  

\section{Molecular hydrogen content}\label{H2_surf_dens_map}
Nowhere in this work have we made provisions in the star formation models for a possible molecular component of each galaxy's ISM.  Stars are known to form from molecular clouds, and so the presence of molecular gas is expected for each galaxy.  Despite this expectation, however, no published CO detections exist for NGC~2915 or NGC~1705.  In this work we therefore could not \emph{directly} study the possible links between the star-forming properties of the galaxies and their \emph{molecular} inter-stellar media.  

\citet{bigiel_2008} showed that a Schmidt-type power law with index $N=1.0\pm2$ relates \sfrsd\ to the molecular gas surface density (\htwosd) across a sample of THINGS spiral galaxies, implying that H$_2$ forms stars at a constant efficiency.  This relation allows an estimate of the molecular content of a system to be obtained from its observed star formation activity.  Using the total star formation rate surface density maps of NGC~2915 and NGC~1705, respective total molecular gas masses of $M_{H_2}=5.7^{+3.4}_{-2.1}\times 10^7$~\msun\ and $M_{H_2}=7.4^{+4.3}_{-7.0}\times 10^7$~\msun\ are inferred to be contained within each galaxy's $R_{25}$ isophotal radius.  The \hi masses of NGC~2915 and NGC~1705 are known to be $\sim 5.5\times 10^8$~\msun\ \citep{elson_2010a_temp} and $\sim 1.1\times 10^8$~\msun\ \citep{meurer_1705_2}, respectively, suggesting molecular-to-atomic-\hi mass ratios of $M_{H2}/M_{HI}=0.08$ and $M_{H2}/M_{HI}=0.48$.  These mass ratios are lower than the average ratio of $M_{H2}/M_{HI}=0.89$ for the 23 gas-rich late-type galaxies from the THINGS sample of \citet{leroy_THINGS}.  This is not surprising given the small sizes of the stellar disks of NGC~2915 and NGC~1705 relative to the sizes of their extended \hi disks.  

\begin{figure}
	\begin{centering}
	\includegraphics[angle=0,width=1\columnwidth]{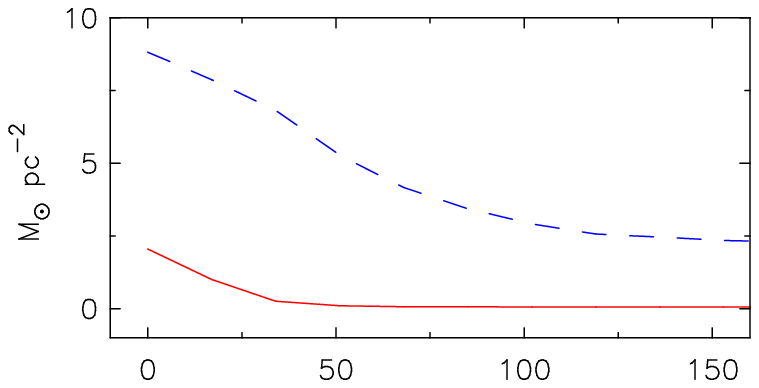}
	\includegraphics[angle=0,width=1\columnwidth]{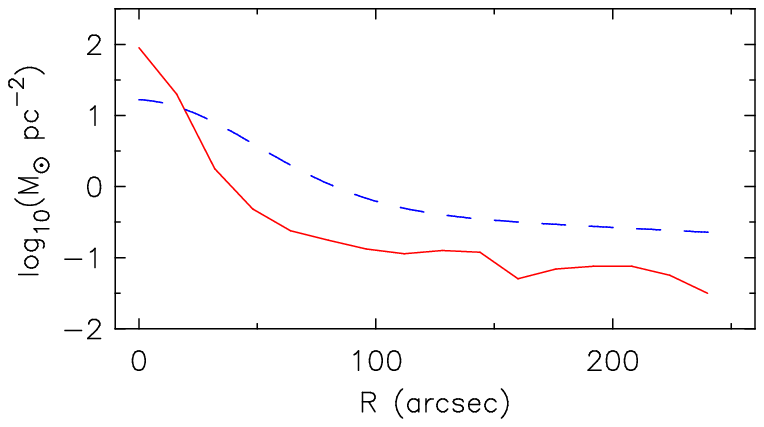}

	\caption{Observed \hi and inferred H$_2$ radial profiles (blue-dashed and solid red curves, respectively) for NGC~2915 and NGC~1705 (top and bottom panels, respectively).  Note the logarithmic ordinate axis of the bottom panel.  The \hi radial profiles for NCG~2915 and NGC~1705 are taken from \citet{elson_2010a_temp} and Elson~et~al. (2012, in prep.), respectively.}
	\label{2915_1705_H2_profiles}
	\end{centering}
\end{figure}

Radial profiles of the SFR surface density maps were used together with the above-mentioned Schmidt-type power law from \citet{bigiel_2008} to generate the H$_2$ surface density radial profiles shown in Fig.~\ref{2915_1705_H2_profiles}.  For comparative purposes each galaxy's \hi profile is also shown.  The inferred H$_2$ surface densities within the $R_{25}$ radius of NGC~2915's stellar disk are much lower ($\lesssim 2$~\msun~pc$^{-2}$) than the corresponding \hi surface densities.   Incorporating them into the star formation laws would result in an increase of the gas surface density by a factor of 1.2 at most.  Such an increase will not significantly affect the results presented in this work.  From the \qgas\ radial profiles of NGC~2915 and NGC~1705 presented in Figs.~\ref{2915_toomre_map} and \ref{1705_toomre_map} it is clear that a boost factor of at least $\sim 2$ is required to yield the inner most portions of each galaxy's gaseous disk gravitationally unstable.  

Because of the intense star burst activity near the centre of NGC~1705, some of the system's inferred H$_2$ surface densities are much higher than the corresponding \hi surface densities.  Caution has to be exercised in the case of NGC~1705, however.  Although the THINGS sample of late-type galaxies of \citet{bigiel_2008} does include 11 late-type dwarfs, none of those systems exhibit star formation activity as extreme as that of NGC~1705.  It is highly likely that extreme systems such as NGC~1705 do not obey the same star formation law as do more typical star forming systems.  In this sense we cannot quantitatively comment on the extent to which an incorporation of NGC~1705's H$_2$ content into the star formation models will affect the results presented in this work.  

\section{Summary}\label{summary}
This paper deals with the investigations of various star formation models for the blue compact dwarf galaxies NGC~2915 and NGC~1705.  Both galaxies contain small stellar disks embedded in much larger \hi disks.  We have used GALEX far ultra-violet and \emph{Spitzer} 24~\micron\ imaging of each galaxy to estimate its total star formation rate.  New deep, high-resolution \hi synthesis observations of each system have been used to quantify the distribution and kinematics of its neutral ISM.  

The star formation models incorporate various properties of each galaxy's ISM to determine which regions within the galaxy should be unstable against the tendency to gravitationally collapse and hence form high-mass stars on varying galactic length scales.  The success or failure of these models sheds light on the interplay between the key properties of the ISM that regulate the star formation activity.  The star formation thresholds in NGC~2915 and NGC~1705 are not purely local phenomena.  The Toomre~single-fluid $Q_{gas}$ criterion as well as the stars+gas and DM+gas two-fluid criteria ($Q_{gas,*}$ and $Q_{gas,DM}$ respectively) all use the epicyclic frequency at a given galactocentric radius within the galaxy as a measure of its kinematics.  The \qgas\ criterion incorrectly predicts the entire \hi disks of both NGC~2915 and NGC~1705 to be sub-critical.  This result is consistent with that of \citet{leroy_THINGS} who found almost no area of the inner disk for their sample of 23 THINGS galaxies to be formally unstable to gravitational collapse in a Toomre \qgas\ context.  These results suggest the Toomre \qgas\ criterion to be of limited utility in terms of distinguishing star-forming regions of high and low efficiency.

The stellar potential of each galaxy is expected to play a crucial role in regulating the star formation activity.  Despite this expectation, however, the \QGS\ criterion also fails for NGC~2915 whose entire \hi disk is predicted to be sub-critical.  The model cannot describe NGC~2915's central star formation activity.  The situation improves for NGC~1705 for which perturbation length scales of $\lambda\sim 200$~pc lead to predictions of its inner \hi disk being formally gravitationally unstable.  

Disk dark matter can account for the observed star formation activity in NGC~2915 and NGC~1705.  Both galaxies are known to be extremely dark-matter-dominated.  The fraction of their dark matter that is co-located within the \hi disk is treated as contributing to the self-gravity of the ISM.  Assuming a pseudo isothermal sphere parameterisation of each galaxy's dark matter halo, both NGC~2915 and NGC~1705 have formally unstable inner \hi disks.  This result does not hold  for the case in which the distribution of the assumed disk dark matter exactly traces the \hi distribution.  

The final star formation model investigated in this paper is a shear-based model built on the premise that it is the time available for clouds to collapse in the presence of rotational shear that regulates a galaxy's star formation.  Indeed, adopting the rotational shear as a characterisation of the global kinematics allows the \sgas\ criterion to accurately locate the unstable parts of both NGC~2915's and NGC~1705's \hi disks without the need for stellar or dark matter self-gravity.  The same general result was obtained by \citet{leroy_THINGS} who found the inner disks of their galaxies to be more nearly super-critical in the context of a shear-based \sgas\ criterion than in the context of a \qgas\ criterion.

In conclusion, despite NGC~2915 and NGC~1705 both being unusual star-forming galaxies, their star-formation activity can be partly understood in terms of the self-gravity of the ISM which, in turn, is controlled by the combined effects of the gravitational potentials of their various mass components.  Alternatively, the star formation activity in each system can be accounted for by correctly characterising its global kinematics.  Throughout this work, in lieu of published CO data, we have ignored any possible H$_2$ components.  Molecular hydrogen must be present, it is unlikely that the observed star formation could be taking place without it.  Based on our measured star formation rates, we have estimated the amount of H$_2$ within each galaxy to be of the order of $\sim 10^7$~\msun\ ($\sim2-3$ orders of magnitude less than the dynamical mass of each galaxy).  Although we cannot reliably incorporate the effects of this mass component into our star formation models, they are not expected to significantly affect the results.

\section{Acknowledgements}\label{acknowledgements}
The work of ECE is based upon research generously supported by the South African SKA project.  All authors acknowledge funding received from the South African National Research Foundation.  The work of WJGdeB is based upon research supported by the South African Research Chairs Initiative of the Department of Science and Technology and the National Research Foundation.  The Australia Telescope Compact Array is part of the Australia Telescope which is funded by the Commonwealth of Australia for operation as a National Facility managed by CSIRO.   This work is based [in part] on observations made with the \emph{Spitzer} Space Telescope and the Galaxy Evolution Explorer, which are operated by the Jet Propulsion Laboratory, California Institute of Technology under contracts with NASA. Finally, all authors thank the anonymous referee for constructive comments that improved the quality of the paper.

\end{document}